\documentclass[reprint,superscriptaddress,amsmath, amssymb, aps, prb]{revtex4-1}

\usepackage{graphicx}
\usepackage{dcolumn}
\usepackage{bm}
\usepackage{subfigure}
\usepackage{multirow}
\setcitestyle{super,open={},close={}}

\begin{document}

\title{Phase diagram and optimal switching induced by spin Hall effect
\\
in a perpendicular magnetic layer }
\author{Shu Yan}
\email{syan@physics.sc.edu}
\author{Ya.\ B. Bazaliy}
\email{yar@physics.sc.edu}
\affiliation{Department of Physics and Astronomy, University of South Carolina, Columbia, South Carolina 29208}
\date{\today}

\begin{abstract}
In a ferromagnet/heavy-metal bilayer device with strong spin Hall effect an in-plane current excites magnetic dynamics through spin torque. We analyze bilayers with perpendicular magnetization and calculate three-dimensional phase diagrams describing switching by external magnetic field at a fixed current. We then concentrate on the case of a field applied in the plane formed by the film normal and the current direction. Here we analytically study the evolution of both the conventional ``up''/``down'' magnetic equilibria and the additional equilibria created by the spin torque. Expressions for the stability regions of all equilibria are derived, and the nature of switching at each critical boundary is discussed. The qualitative picture obtained this way predicts complex hysteresis patterns that should occur in bilayers. By analyzing the phase portraits of the system we show that when the spin torque induced equilibrium exists, switching between ``up'' and ``down'' states proceeds through it as an intermediate state. Using numeric simulations we analyze the switching time and compare it to that of a conventional spin torque device with collinear magnetizations of the polarizer and the free layer.
\end{abstract}

\pacs{Valid PACS appear here}

\maketitle

\section{INTRODUCTION}

Recently a number of investigations focused on bilayer structures consisting of a ferromagnetic (F) layer and a nonmagnetic (N) layer with strong spin-orbit (SO) interaction made of heavy-metals such as Pt, Ta or W.\cite{PhysRevLett.101.036601, miron2010current, pi2010tilting, PhysRevLett.106.036601, Miron2011, PhysRevLett.107.146602, PhysRevLett.109.096602, liu2012spin, pai2012spin, kim2012layer} It was theoretically predicted and experimentally observed that when an in-plane electric current is being applied, the itinerant electrons inside the nonmagnetic layers become spin polarized due to the strong spin-orbit coupling and exert a spin torque on the ferromagnetic layers. This additional torque contributes to the magnetization dynamics described by the Landau-Lifshitz-Gilbert (LLG) equation. Up to now two different models have been proposed to account for the effects. One of them \cite{miron2010current, Miron2011} treats the bilayer structure as a two-dimensional system with strong interfacial Rashba spin-orbital coupling due to the structural inversion symmetry breaking in the direction normal to the interface.\cite{bychkov1984oscillatory} This model leads to a field-like torque directed along $\hat{\bf{m}}\times ({\bf{j}}_e \times \hat{\bf{z}})$,\cite{PhysRevB.78.212405, PhysRevB.79.094422, matos2009spin, gambardella2011current, PhysRevB.88.214417} where $\hat{\bf{m}} = {{\bf{M}}/M_s}$ is the magnetization unit vector ($M_s$ represents the constant absolute value of the magnetization ${\bf{M}}$), ${\bf{j}}_e$ is the in-plane electric current density, and the interface normal vector is pointing along the $\hat{\bf{z}}$ direction. The other model \cite{PhysRevLett.101.036601, PhysRevLett.106.036601, PhysRevLett.107.146602, PhysRevLett.109.096602, liu2012spin, pai2012spin} is based on the interfacial diffusion of the pure spin current that originates in the heavy-metal layers due to the bulk spin Hall effect (SHE)\cite{PhysRevLett.83.1834, PhysRevLett.85.393, PhysRevLett.92.126603, jungwirth2012spin} and leads to spin transfer torque (STT) dynamics \cite{slonczewski1996current, ralph2008spin} in the magnetic layers. In the SHE model the torque is directed along $\hat{\bf{m}}\times [\hat{\bf{m}}\times ({\bf{j}}_e \times \hat{\bf{z}})]$.\cite{PhysRevB.87.174411} This type of torque is called a Slonczewski, or damping-like, or adiabatic torque in the literature.

Several experiments showed that an in-plane electric current flowing through the structure is able to switch the magnetization of the ferromagnetic layer.\cite{Miron2011, PhysRevLett.109.096602, liu2012spin,pai2012spin} In those experiments the F layers were magnetized perpendicular to the film plane. It is believed that the observed magnetic reversal can only be induced by the damping-like torque. The reasons for this are (a) experimentally measured switching phase diagrams are in accord with the macrospin model calculations\cite{PhysRevLett.109.096602,PhysRevB.89.104421}, and (b) due to its symmetry, a field-like torque, if there is any, does not favor either ``up'' or ``down'' state of the perpendicular magnetization, and therefore should not contribute to switching. These arguments seem to favor the SHE-based model, however, subsequent calculations \cite{PhysRevB.85.180404, PhysRevB.86.014416, PhysRevLett.108.117201} suggested that the model based on Rashba coupling generates both field-like and damping-like torques, and thus is also capable of describing the switching.

\begin{figure}[b]
    \centering
    \includegraphics[scale=0.6]{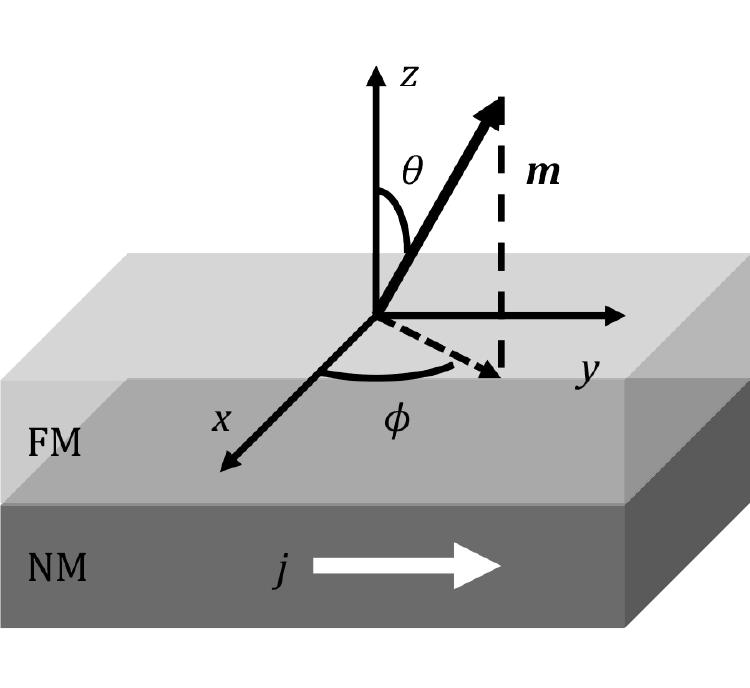}
    \caption{Schematic diagram of the bilayer spin Hall effect device. Electric current $j$ flows along the y axis.}
    \label{fig:device}
\end{figure}

Despite the fact that the underlying torque mechanism is still not fully understood, a thorough study of the switching behavior based on the existing experimental results is of importance for analysis and prediction. In this paper we study magnetic switching induced by an externally applied magnetic field $\bf H$ at a fixed in-plane electric current. It is assumed that the magnetic anisotropy energy of the F layer is uniaxial
\begin{equation}\label{eq:anisotropy}
    E({\bf m}) = -K(\hat{\bf{m}} \cdot \hat{\bf{z}})^2.
\end{equation}
Magnetization is switched between ``up'' and ``down'' at critical fields ${\bf H}_c$ that form a surface in the three-dimensional $H$-space. Without electric current and for the magnetic anisotropy given by (\ref{eq:anisotropy}) this surface is an axially symmetric figure of revolution with a cross-section given by the astroid curve.\cite{stoner1948mechanism} The presence of the in-plane current ${\bf j}_e$ breaks the axial symmetry of the critical surface ${\bf H}_c$. To find its shape we use the method of Refs.~\onlinecite{PhysRevB.61.12221, PhysRevB.88.054408}. A related problem of finding the critical field magnitude as a function of current $H_c(j_e)$ for a given field direction was considered numerically in a number of publicatoins.\cite{PhysRevB.83.054425} In our treatment we consider only the damping-like torque. Since the field-like torque can be compensated by an external field, its presence simply shifts the critical surface and we do not include it in the calculations. First, we find a three-dimensional critical surface. Second, we study a particular section of this surface with ${\bf H}$ confined to the plane formed by the directions of current ${\bf{j}}_e$ and magnetic easy axis. Here we derive analytic formulas for the switching boundaries ${\bf H}_c$. We also observe that, in accord with numeric investigations,\cite{lee2013threshold} in the presence of a sufficiently large current an extra equilibrium direction appears in addition to the ``up" and ``down" equilibria, and get analytic expression for its location. By analyzing the phase portraits of the system we show that if additional equilibrium exists, switching form ``up'' to `` down'' state proceeds through this intermediate state. Using numeric simulations we analyze the switching time and compare it to that of a perpendicular polarizer device.

\section{GENERAL DESCRIPTION OF THE THEORETICAL APPROACH}\label{general}

Magnetization dynamics of the ferromagnetic layer in the macrospin approximation is governed by the LLG equation:
\begin{equation} \label{LLG}
    \frac{d{\bf{M}}}{dt} = -\gamma \mu_0 ({\bf{M}}\times{{\bf{H}}}_{\text{eff}})+
    \frac{\alpha}{M_s}\left({\bf{M}} \times \frac{d{\bf{M}}}{dt} \right),
\end{equation}
where $\alpha $ is the Gilbert damping factor, $\gamma$ is the gyromagnetic ratio, and ${\bf{H}}_{\text{eff}}$ is the total effective field. The standard LLG equation can be transformed into
\begin{equation} \label{LLG_eq}
\frac{d\hat{\bf{m}}}{dt}=-\hat{\bf{m}}\times{\bf{h}}_{\text{eff}}-\alpha\hat{\bf{m}}
\times\hat{\bf{m}}\times{\bf{h}}_{\text{eff}},
\end{equation}
where the field is rescaled as ${\bf{h}}_{\text{eff}}={\bf{H}}_{\text{eff}}/H_k$  using the characteristic anisotropy field $H_k = 2K/ \mu_0 M_s$, and the time is rescaled as $t\to t'=\gamma \mu_0 H_k t/(1+\alpha ^2)$. Hereafter, all the field-related terms that are written in lowercase letters are dimensionless (normalized by $H_k$).

The method \cite{PhysRevB.61.12221, PhysRevB.88.054408} can be summarized as follows. A stationary solution $\hat{\bf{m}}_0$ of Eq.\ (\ref{LLG_eq}) satisfies the equilibrium condition $\hat{\bf{m}}\times{\bf{h}}_{\text{eff}}=0$, which indicates that the magnetization at equilibrium should be parallel to the total effective field, i.e.,  ${\bf{h}}_{\text{eff}} = \lambda \hat{\bf{m}}_0$ with arbitrary $\lambda$. Total effective field is given by ${\bf{h}}_{\text{eff}}={\bf{h}}-\nabla \varepsilon +{\bf{h}}_{\text{sp}}$, where ${\bf{h}}$ is the external field, $\varepsilon = E/(\mu_0 H_k M_s)$ is the rescaled anisotropy energy, and ${\bf{h}}_{\text{sp}}$ is the spin torque effective field
\begin{equation}\label{eq:hsp_expression}
    {\bf h}_{\text{sp}} = \alpha_j [{\bf m} \times ({\bf e}_j \times \hat z)],
\end{equation}
where ${\bf e}_j$ is a unit vector in the electric current direction and $\alpha_j$ is a spin torque strength parameter, proportional to the electric current density. Equation ${\bf{h}}_{\text{eff}} = \lambda \hat{\bf{m}}_0$ can be solved for the external field as ${\bf h} = \lambda {\bf m}_0 +\nabla\varepsilon({\bf m}_0) - {\bf h}_{sp}({\bf m}_0)$. The meaning of this formula is that for any given magnetization direction there is a whole line of external fields, parameterized by $\lambda$, for all of which this direction is an equilibrium (stable or unstable). In spherical coordinates with three orthogonal unit vectors defined as $\hat{\bf{m}}=\sin\theta \cos\phi {\hat{\bf{x}}} + \sin\theta \sin\phi {\hat{\bf{y}}} + \cos\theta {\hat{\bf{z}}}$, $\hat{\bf{\theta}}=\partial\hat{\bf{m}}/\partial\theta$, and $\hat{\bf{\phi}}=(1/\sin\theta)\partial\hat{\bf{m}}/\partial\phi$, we get
\begin{equation} \label{h_ext}
    {\bf{h}}=\lambda \hat{\bf{m}}_0+\left(\partial _{\theta }\varepsilon -h_{\text{sp}}^{\theta }\right)_0 \hat{\bf{\theta}}_0+\left(\frac{1}{\sin  \theta
    }\partial _{\phi }\varepsilon -h_{\text{sp}}^{\phi }\right)_0 \hat{\bf{\phi}}_0 \ ,
\end{equation}
where $\partial_\theta$ stands for $\partial/\partial\theta$ and the superscript $\theta $ indicates the $\hat{\bf{\theta}} $ component of a vector field (e.g., $h_{\text{eff}}^\theta = \bf{h}_{\text{eff}} \cdot \hat{\bf{\theta}} $), etc. Equation (\ref{h_ext}) maps the 3D space $\{\lambda,\theta_0,\phi_0\}$ to the 3D field space $\{h_x, h_y, h_z\}$.

Next, stability of equilibrium states is analyzed. This is done by expanding Eq.\ (\ref{LLG_eq}) in small deviations $\hat{\bf{m}} = \hat{\bf{m}}_0 + \delta \hat{\bf{m}}$ up to the linear terms.  such an expansion produces two coupled linear differential equations
\begin{equation}
    \left(
    \begin{array}{c}
        \dot{\delta \theta } \\
        \sin  \theta _0\dot{\delta \phi } \\
    \end{array}
    \right)=A\left(\theta _0,\phi _0\right)\left(
    \begin{array}{c}
        \delta \theta  \\
        \sin  \theta _0\delta \phi  \\
    \end{array}
    \right),
\end{equation}
with matrix $A(\theta_0,\phi_0)$ given by
\begin{equation}
    A =
    \begin{bmatrix}
        \partial _{\theta }(\alpha  h_{\text{eff}}^{\theta }+h_{\text{eff}}^{\phi }) & \frac{1}{\sin  \theta}\partial _{\phi }(\alpha h_{\text{eff}}^{\theta }+h_{\text{eff}}^{\phi }) \\[0.3em]
        \partial _{\theta }(\alpha  h_{\text{eff}}^{\phi }-h_{\text{eff}}^{\theta }) & \frac{1}{\sin  \theta}\partial _{\phi }(\alpha h_{\text{eff}}^{\phi }-h_{\text{eff}}^{\theta })
    \end{bmatrix}
    \label{eq:matrixA}
\end{equation}
Stationary solutions can be classified as stable or unstable using the eigenvalues $\mu_\pm $ of the matrix $A$, which are uniquely determined by its determinant, $\text{det} A$, and trace, $\text{tr} A$.\cite{hirsch2004differential} An equilibrium is stable when both eigenvalues $\mu_\pm $ are either complex numbers with negative real parts, or negative real numbers, which leads to the stability criteria
\begin{equation} \label{criteria}
    \text{det} A>0 \text{ and } \text{tr} A<0 \ ,
\end{equation}
that are applied to select the subset of the line ${\bf h}(\lambda,\theta_0,\phi_0)$ for which $(\theta_0,\phi_0 )$ equilibrium is a stable, i.e., to find the values of $\lambda$ for which conditions (\ref{criteria}) are satisfied. By evaluating expression (\ref{eq:matrixA}) at external field specified by Eq.~(\ref{h_ext}) one obtains $A(\lambda)$. We find that for the arbitrary form of spin torque and arbitrary anisotropy energy, $\text{tr} A(\lambda)$ is linear in $\lambda$ with a negative coefficient, and $\text{det} A(\lambda)$ is a quadratic function of $\lambda$ with a positive quadratic coefficient. To simplify the expressions, we introduce a vector field $\bf{f}=-\nabla \varepsilon +{\bf{h}}_{\text{sp}}$ and its gradient
\begin{equation}\label{vecfld}
    \nabla \bf{f}=\left[
    \begin{array}{cc}
        \partial_\theta f^\theta & \partial_\theta f^\phi \\
        \partial_\phi f^\theta & \partial_\phi f^\phi \\
    \end{array}
    \right]
\end{equation}
(see Appendix \ref{mtxelms} for the explicit expressions). The roots of equations $\text{tr} A(\lambda)=0$ and $\text{det} A(\lambda)=0$ can then be respectively calculated as
\begin{equation}\label{lambda_T_general}
    \lambda_{\text{T}}(\theta_0,\phi_0)=\frac{1}{2}\left[\partial_\theta f^\theta+\partial_\phi f^\phi + \frac{1}{\alpha}\left(\partial_\theta f^\phi - \partial_\phi f^\theta\right)\right],
\end{equation}
\begin{eqnarray} \label{lambda_pm_general}
    \lambda_{\pm}(\theta_0,\phi_0) &=& \frac{\partial_\theta f^\theta+\partial_\phi f^\phi}{2}
    \\
    \nonumber
    && \pm \sqrt{\left(\frac{\partial_\theta f^\theta - \partial_\phi f^\phi}{2}\right)^2 + \partial_\theta f^\phi \partial_\phi f^\theta}.
\end{eqnarray}
In terms of the critical values $\lambda_{\text{T}}$ and $\lambda_{\pm}$, the stability criteria become
\begin{equation}\label{stb}
    \begin{cases}
        \lambda > \text{Max}(\lambda_\text{T},\lambda_{+}) & \text{if } \lambda_\text{T} \geqslant \lambda_{-}, \\
        \lambda_\text{T} < \lambda < \lambda_{-} \text{ or } \lambda > \lambda_{+} & \text{if } \lambda_\text{T} < \lambda_{-}.
    \end{cases}
\end{equation}
When $\lambda_\pm$ are complex, $\text{det} A$ is always positive and criteria (\ref{stb}) can be further simplified as $\lambda > \lambda_{\text{T}}$. The full classification is given in Appendix \ref{table}.

Substituting functions $\lambda_{+}$, $\lambda_{-}$ or $\lambda_\text{T}$ for $\lambda$ in Eq.\ (\ref{h_ext}) one generates three surfaces in the field space, which we denote as $S_{+}$, $S_{-}$ and $S_\text{T}$ respectively. Their physical meaning is that at least one equilibrium changes its stability when external field crosses such a surface. The equilibrium is either locally destabilized when the $S_\text{T}$ surface is crossed, or merges with a saddle point when the  $S_\pm$ surfaces are crossed.\cite{PhysRevB.84.064422} The entire critical surface $S$ is constructed from the parts of $S_{+}$, $S_{-}$ and $S_\text{T}$ as explained in Ref.~\onlinecite{PhysRevB.88.054408}.

\section{3D PHASE DIAGRAM}\label{sec:3D}

\begin{figure}[t]
    \centering
    \includegraphics[scale=0.4]{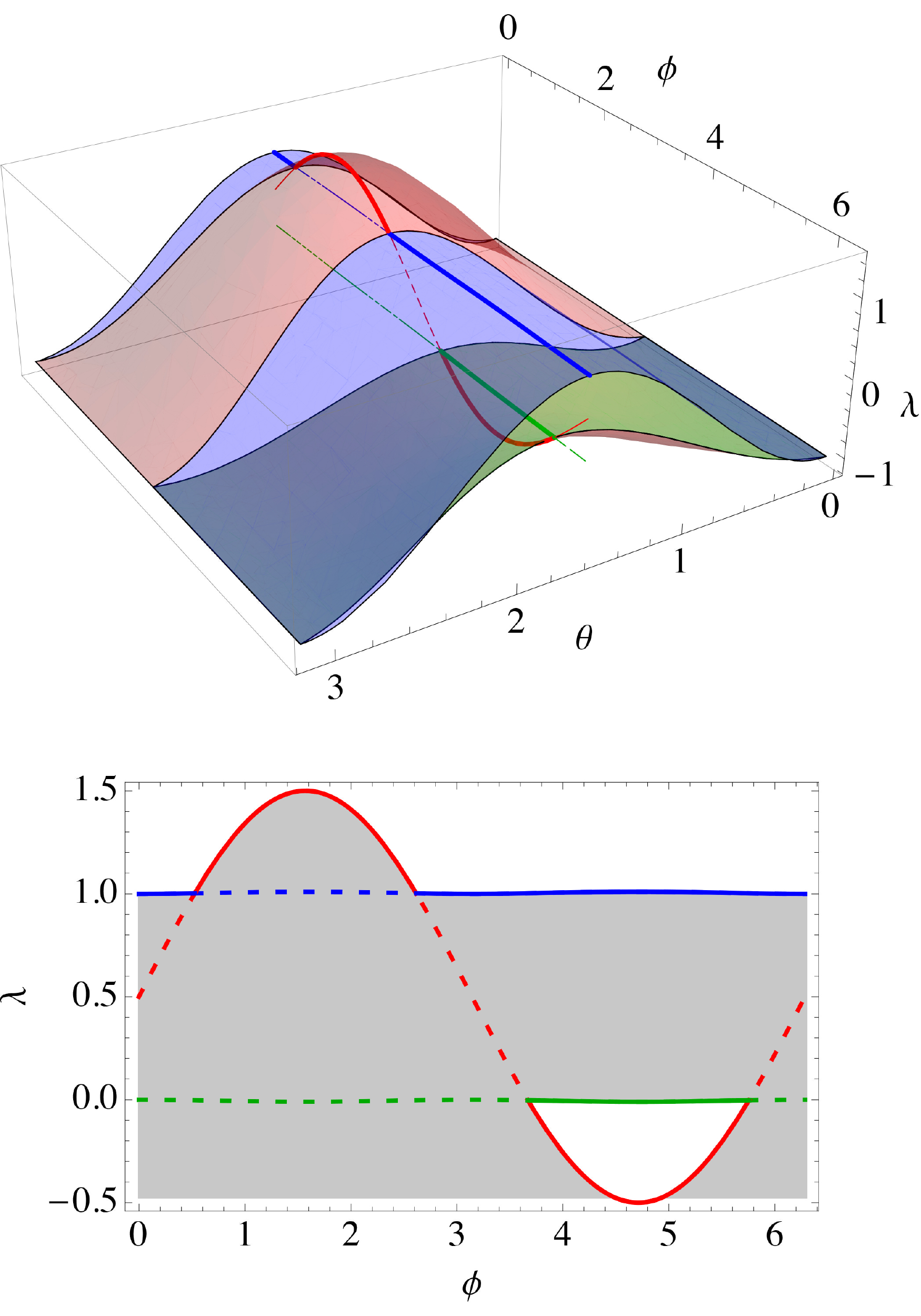}
    \caption{(color online) The critical $\lambda$ surfaces (upper panel) and the cross-section at $\theta = \pi/2 $ (lower panel). The $\lambda_{+}$, $\lambda_{-}$ and the $\lambda_{\text{T}}$ surfaces are plotted in blue, green and red, respectively. The $\lambda_{-}$ surface is always below $\lambda_{+}$. The $\lambda_{\text{T}}$ surface in this case is separated into two parts, one above $\lambda_{+}$ and the other below $\lambda_{-}$. In the lower panel, stable and unstable regions are denoted by white and gray, separated by the active parts of the three lines. The damping parameter and the spin torque strength are set to be $\alpha = 0.1$ and $\alpha_\text{j} = 0.1$.}
    \label{fig:lambda}
\end{figure}

In this section we construct the three-dimensional critical surface using the method of Sec.~\ref{general}. The dimensionless perpendicular anisotropy energy has the form $\varepsilon = -\cos ^ 2 \theta/2$. We set the in-plane current to be along the $+\hat{\bf{y}}$ direction, and the current induced field is then ${\bf{h}}_{\text{sp}} = \alpha_j \hat{\bf{m}} \times \hat{\bf{x}}$ with $\alpha_j$ given by $\alpha_j=\theta_\text{SH}j_\text{e}/j_0$, where $\theta_\text{SH}$ is the spin Hall angle, $j_0=2e M_\text{s} d_\text{F}H_k/\hbar$ is the characteristic current density, and $d_\text{F}$ is the thickness of the F layer. For brevity, we drop index ``0'' for the equilibrium direction. The critical values of $\lambda$ calculated according to (\ref{lambda_T_general}) and (\ref{lambda_pm_general}) specialize to
\begin{equation}\label{l_T}
    \lambda_{\text{T}}(\theta,\phi) = \frac{\sin^2 \theta}{2} - \cos^2 \theta - \frac{\alpha_\text{j}}{\alpha}\sin \theta \cos \phi,
\end{equation}
\begin{equation}\label{l_pm}
    \lambda_{\pm}(\theta,\phi) = \frac{\sin^2 \theta}{2} - \cos^2 \theta \pm \sin \theta \sqrt{\frac{\sin^2\theta}{4} - \alpha_j^2\cos^2\phi}.
\end{equation}
For $\alpha_j = 0$ one finds $\lambda_\text{T}$ to be in the midpoint of the interval $ (\lambda_{-}$, $\lambda_{+})$ for any direction $(\theta,\phi )$. Then, according to the criteria (\ref{stb}), $S_+$ constitutes the entire critical surface. For non-zero current $\lambda_\text{T}$ may leave the interval $ (\lambda_{-}$, $\lambda_{+})$ for some values of $(\theta,\phi)$. In those cases switching boundaries should be selected for every direction individually. Fig.\ \ref{fig:lambda} shows the three critical values as functions of equilibrium angles at nonzero current with $\alpha_j=0.1$ and $\alpha = 0.1$.

\begin{figure}[t]
    \centering
    \includegraphics[scale=0.4]{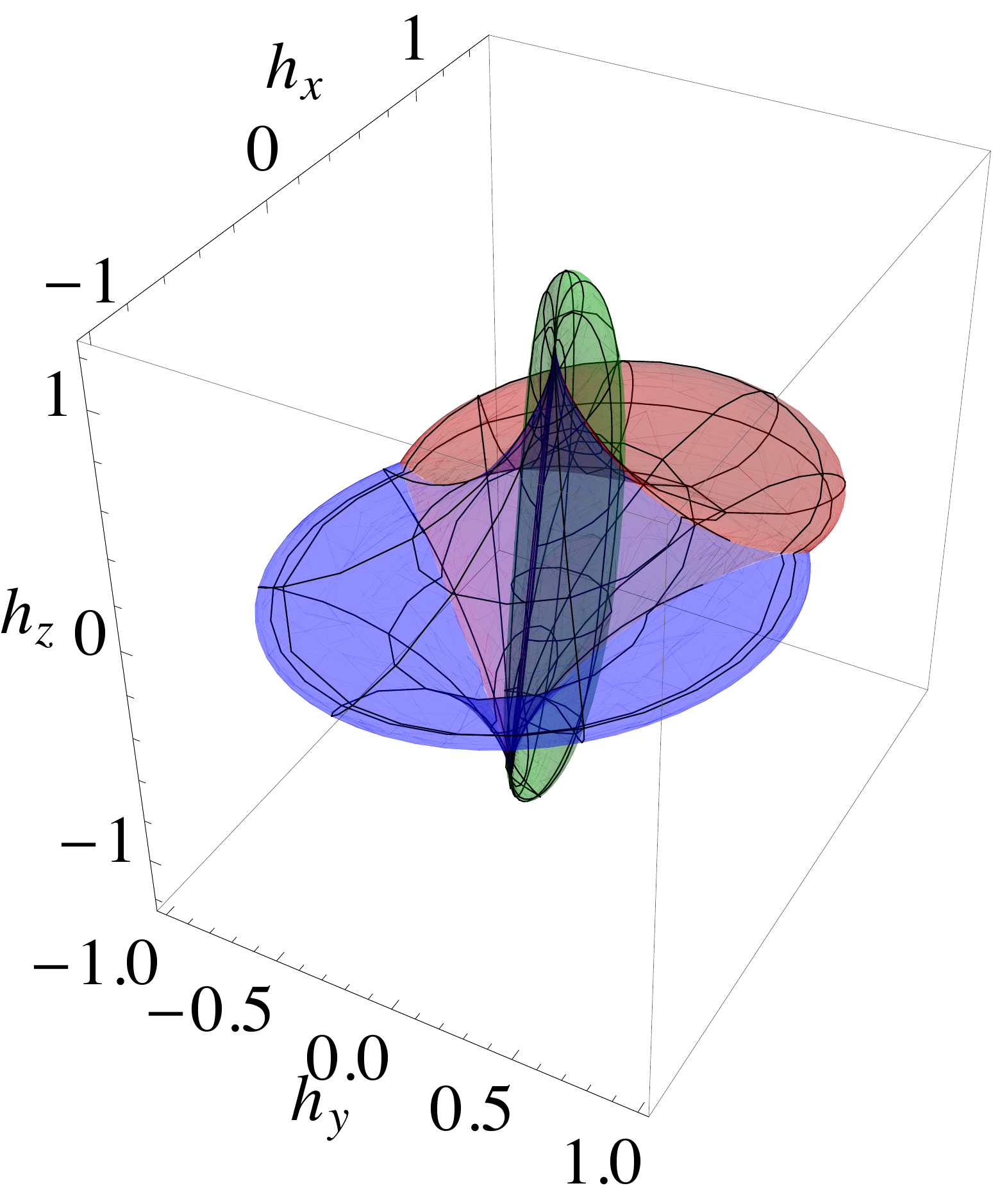}
    \caption{(color online) The critical surface in three-dimensional field space. The same color convention and parameters as in Fig.\ \ref{fig:lambda} are adopted.}
    \label{fig:field}
\end{figure}

For practical calculations we decompose Eq.\ (\ref{h_ext}) into Cartesian coordinates
\begin{subequations}
    \begin{align}
        \label{h_x}
        & h_x = (\lambda + \cos^2\theta) \sin\theta \cos\phi, \\
        \label{h_y}
        & h_y = (\lambda + \cos^2\theta) \sin\theta \sin\phi - \alpha_j \cos\theta, \\
        \label{h_z}
        & h_z = (\lambda - \sin^2\theta) \cos\theta + \alpha_j \sin\theta \sin\phi.
    \end{align}
\end{subequations}
For arbitrary $\lambda$'s these equations represent the mapping of the $(\lambda,\theta,\phi)$ space to the $(h_x, h_y, h_z)$ space for the case of uniaxial anisotropy and chosen electric current direction. If functions $\lambda_T(\theta,\phi)$ (\ref{l_T}) or $\lambda_{\pm}(\theta,\phi)$  (\ref{l_pm}) are substituted for $\lambda$, one obtains parametric expressions for the critical surfaces $S_T$ and $S_{\pm}$ with parameters $(\theta, \phi)$ running through all possible values, $0 \leqslant \theta \leqslant \pi$ and $0 \leqslant \phi \leqslant 2\pi$. The 3D phase diagram in field space with the same parameters is shown in Fig.\ \ref{fig:field}.

\section{PHASE DIAGRAM IN THE Y-Z PLANE}\label{yzdiagram}

The 3D phase diagram is quite difficult to use due to the complicated shape of the critical surface $S$. The related experiments are often performed with the field being confined within the $y-z$ plane.\cite{PhysRevLett.109.096602,PhysRevB.89.104421} Here we study in detail a section of $S$ corresponding to the external field $\bf h$ confined to such a plane. This section is a line $\bar S$ in the 2D space $(h_y, h_z)$.

A field in the $y-z$ plane satisfies a constraint $h_x = 0$. According to Eq.~(\ref{h_x}) this implies a relationship between $\theta$, $\phi$, and $\lambda$. On the one hand, this relationship allows one to express the equilibrium angles $(\theta,\phi)$ as functions of $(h_y,h_z)$ and study how the equilibria evolve as a function of applied field. On the other hand, Eq.\ (\ref{h_x}) can be used to find the section $\bar S$. While the surface $S$ is given by a parametric formula with independently varying $\theta$ and $\phi$ as explained in Sec.~\ref{sec:3D}, the line $\bar S$ is found from the same formula with $\theta$ and $\phi$ being related to each other by a requirement that Eq.~(\ref{h_x}) holds with $\lambda = \lambda_{T,+,-}(\theta,\phi)$.

\subsection{Evolution of equilibrium states}\label{sec:EvolutionEquilibriumStates}

Equation~(\ref{h_x}) has three types of solutions: (I) $\phi = \pm \pi/2$, (II)  $\lambda = - \cos^2\theta$, and (III) $\theta = 0,\pi $. Since the value of $\phi$ at $\theta = 0,\pi$ is immaterial, type III can be considered as a special case of type I. Thus we focus on the first two cases. For definiteness, assume $\alpha_j > 0$.

Solutions of type I have $\sin\phi = \pm 1$. They are located on the meridian of the unit sphere lying in the $y-z$ plane. We will call them on-meridian states. Eliminating $\lambda$ from Eqs.\ (\ref{h_y}) and (\ref{h_z}) one finds a system of equations for their polar angles
\begin{eqnarray}
\nonumber
&&   \phi = \pm \pi/2 \ ,
 \\
\label{eq:typeIsolutions}
&&    h_y \cos\theta \mp h_z\sin\theta = \pm \sin\theta\cos\theta - \alpha_j.
\end{eqnarray}
Depending on $h_y$, $h_z$ and $\alpha_j$ there can be four, two or zero equilibrium states of type I.

Solutions of type II have $\lambda = - \cos^2\theta$. Equations \ (\ref{h_y}) and (\ref{h_z}) read
\begin{eqnarray*}
    h_y &=& - \alpha_j \cos\theta,
    \\
    h_x &=& - \cos\theta + \alpha_j \sin\theta \sin\phi.
\end{eqnarray*}
Solving them one finds
\begin{eqnarray}
    \label{eq:typeIIsolutions}
    \cos\theta &=& - \frac{h_y}{\alpha_j},
    \\ \nonumber
    \sin\phi &=& \frac{h_z - h_y / \alpha_j}{\sqrt{\alpha_j^2 - h_y^2}}
\end{eqnarray}
with associated requirements $|h_y| \leq \alpha_j$ and $|(h_z - h_y / \alpha_j)/\sqrt{\alpha_j^2 - h_y^2}| \leq 1$ that define their domain of existence. Having $\phi \neq \pm\pi/2$, type II solutions are away from the $y-z$ plane and will be called off-meridian states. They exist as a pair with the same polar angle $\theta$ and azimuthal angles $\phi$ and $\pi - \phi$.

\begin{figure}[t]
    \centering
    \includegraphics[width=0.35\textwidth]{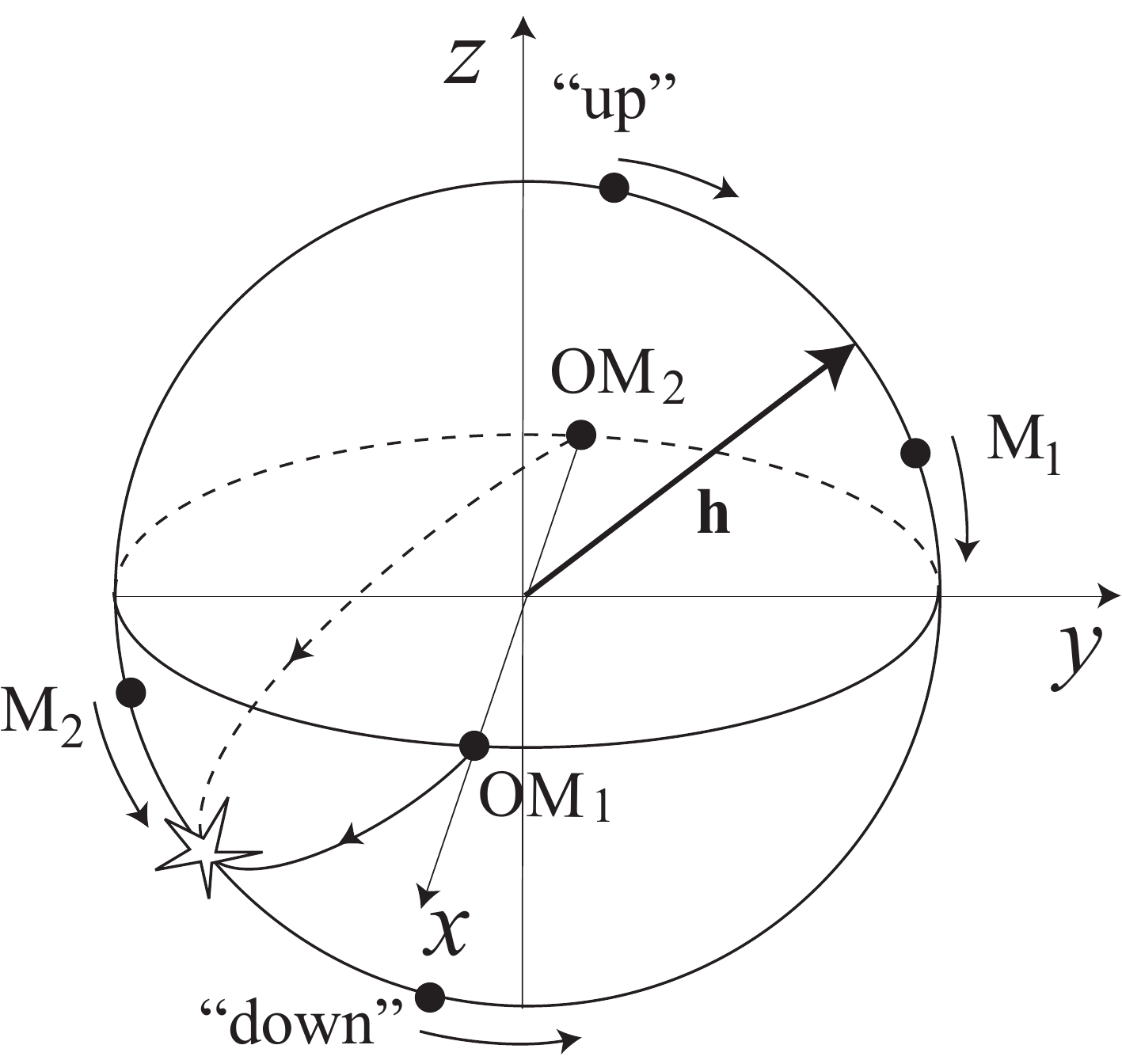}    \caption{Equilibrium states shown as points on the unit sphere for small $\alpha_j > 0$ in the absence of external field. Arrows show the directions of equilibria displacements as an external field $\bf h$ is increased at constant $\alpha_j$. Points $OM_{1}$, $OM_{2}$, and $S_2$ eventually merge into $X$.}
    \label{fig:equilibria_evolution}
\end{figure}

Equilibrium states can be visualized as points on the unit sphere that change their positions when the experimental parameters $\bf h$ and $\alpha_j$ are varied (Fig.~\ref{fig:equilibria_evolution}). In the absence of current and external field the uniaxial magnet exhibits two stable equilibria $\theta = 0,\pi$: the ``up'' and ``down'' states. Due to the axial symmetry of the system the entire equator of the unit sphere forms a circle of unstable equilibrium state.

At non-zero current spin torque breaks the axial symmetry of the problem even in the absence of magnetic field. For $\alpha_j \neq 0$, ${\bf h} = 0$ the continuous set of unstable equilibria along the sphere's equator is reduced to four isolated equilibrium points. Two of them are off-meridian states ($OM_{1,2}$ in Fig.~\ref{fig:equilibria_evolution}) for which Eq.~(\ref{eq:typeIIsolutions}) gives $\theta = \pi/2$ and $\phi = 0, \pi$, i.e., the $\pm\hat x$ directions. The other two are on-meridian states $S_{1,2}$ given by Eq.~(\ref{eq:typeIsolutions}). It will be shown below that they are saddle points. For small $\alpha_j$ the system has six equilibria: slightly displaced ``up'' ($M_{up}$) and ``down'' ($M_{down}$)  on-meridian states, on-meridian states $S_{1,2}$ that are slightly displaced from the equator of the sphere, and the $\pm\hat x$ states $OM_{1,2}$ (Fig.~\ref{fig:equilibria_evolution}). As $\alpha_j$ is increased, the states $S_1$ and $S_2$ approach the ``up'' and ``down'' states respectively. At a critical current they merge pairwise and disappear.

The following useful rules apply to the on-meridian equilibria described by Eq.~(\ref{eq:typeIsolutions}): (1), Increasing current shifts points $M_{up/down}$ clockwise and points $S_{1,2}$ counterclockwise along the meridian; (2), Increasing magnetic field shifts points $M_{up/down}$ along the meridian towards the field directions and points $S_{1,2}$ away from the field direction (see Fig.~\ref{fig:equilibria_evolution}).

Consider now the situation with a small fixed current and a variable external field. For the discussion we will assume a fixed direction of $\bf h$ between $+\hat y$ and $+\hat z$ directions (see Fig.~\ref{fig:equilibria_evolution}). Equations (\ref{eq:typeIIsolutions}) show that as the field magnitude $h$ is increased, the off-meridian states approach the meridian and reach it at a critical field magnitude. Since the two off-meridian states are mirror symmetric with respect to the $y-z$ plane, they reach the meridian simultaneously and merge. Moreover, using Eq.~(\ref{eq:typeIIsolutions}) one can show that the merging point also satisfies Eq.~(\ref{eq:typeIsolutions}), so actually a merging of two off-meridian and one on-meridian equilibrium takes place. This tri-equilibrium merging is not accidental. As discussed in Ref.~\onlinecite{PhysRevB.84.064422}, merging of equilibria has to conserve the winding number and it would be impossible for the two off-meridian equilibria with equal winding numbers to merge without the participation of a third equilibrium with the opposite winding number.

As $h$ is increased further, equilibria $S_2$, $OM_1$, and $OM_2$ merge into one equilibrium $X$ that remains on-meridian. Analysis in the next section shows that $X$ is an unstable focus, analogous to the maximum energy point equilibrium of a conventional (no spin torque) uniaxial magnet subjected the external field. In general, above the critical field the evolution of the four on-meridian equilibria $M_{up}$, $M_{down}$, $S_1$, and $X$ is qualitatively similar to that found at $\alpha_j = 0$. We may conclude that our system has two qualitative regimes: one at low magnetic field where spin torque dominates, and another one at high field where magnetic torque dominates. The spin torque dominated regime is characterized by the presence of two $OM$ equilibria produced by current. In the field dominated regime the current-induced equilibria are gone.

These results are quite natural. The SHE system is equivalent to a conventional spin-transfer device with spin polarizer $\bf p$ directed along $+\hat x$. Spin torque attracts the magnetization to $\bf p$ and repels from $-{\bf p}$. At very large currents spin torque dominates all other torques, so, only two equilibrium points --- one close $\bf p$ and another close to $-{\bf p}$ should remain. In our system we find that it is enough for the spin torque to dominate the magnetic field torque in order to produce these equilibria. This happens because for the easy axis energy (\ref{eq:anisotropy}) and ${\bf p}= \hat x$ the anisotropy torque is equal to zero at ${\bf m} = \pm {\bf p}$.

\subsection{Stability of equilibria analysis and switching phase diagram}
In this section we are going to find the critical line $\bar S$ of equilibrium destabilization. It will be composed from parts produced by type I and type II solutions.

\subsubsection{On-meridian equilibria}
\begin{figure}[t]
    \centering
    \includegraphics[width=0.35\textwidth]{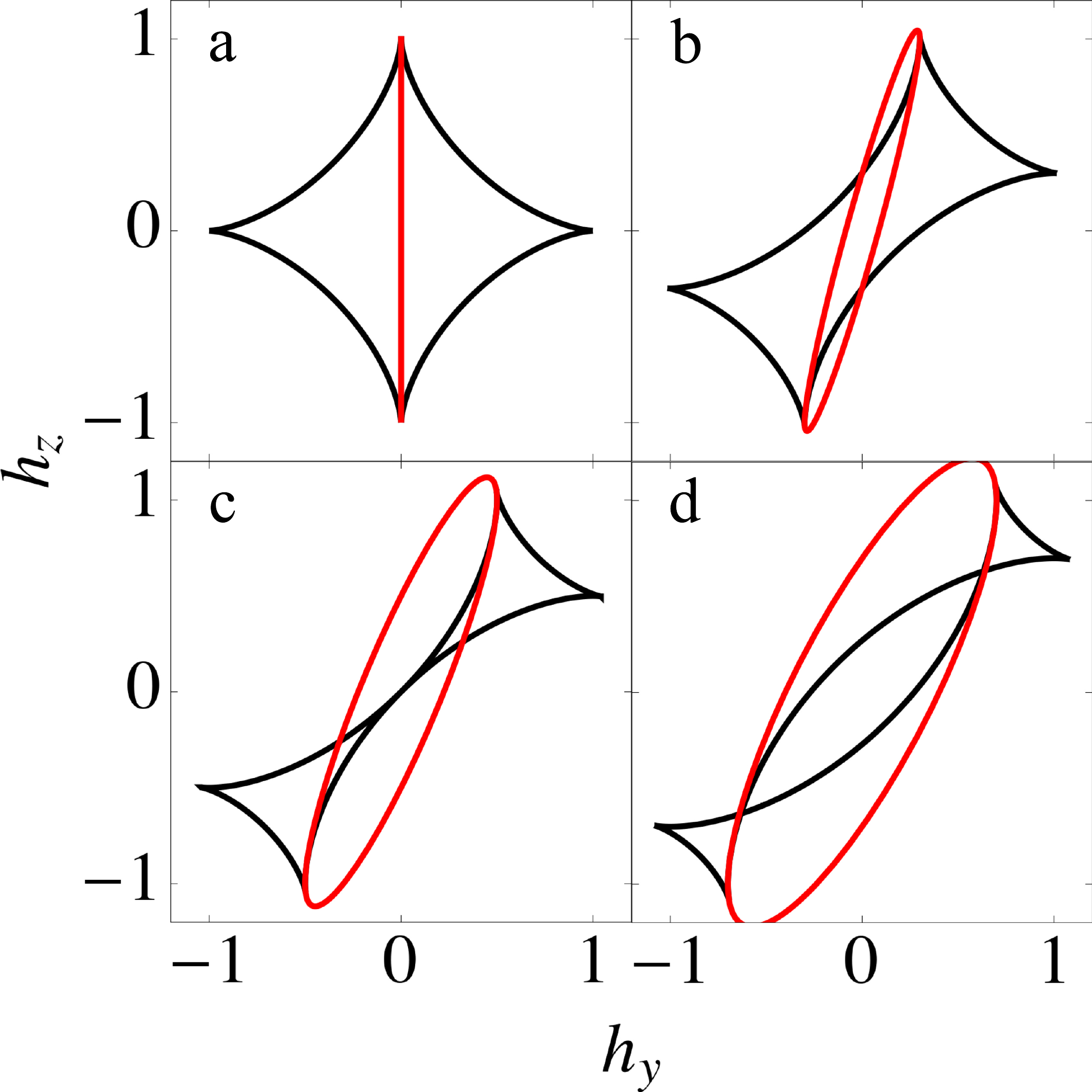}
    \caption{(color online) Lines $\bar S_M$ (black) and the critical field boundary where $OM_1$, $OM_2$ and $S_2$ merge into $X$ (red/grey). The spin torque strength $\alpha_j$ is set to be a. 0.0, b. 0.3, c. 0.5, and d. 0.7.}
    \label{fig_astroid_1}
\end{figure}

Equations (\ref{l_T}) and (\ref{l_pm}) show that for the on-meridian states $\lambda_{\text{T}}$ is the midpoint of $\lambda_\pm$ interval for any current value. Therefore only $\lambda_{+}$ is needed to calculate the critical surface. By substituting $\lambda = \lambda_{+}(\theta,\phi)$ and $\sin\phi = \pm 1$ into Eqs.\ (\ref{h_y}) and (\ref{h_z}), we get an exact parametric form of $\bar S_M$, the line of on-meridian equilibria destabilization. It turns out to be the same as the one found before\cite{PhysRevB.83.054425} using an approximate method.
\begin{equation}\label{astroid}
    \left\{
    \begin{aligned}
        h_y &= \pm \sin^3\theta - \alpha_j \cos\theta \ , \\
        h_z &= - \cos^3\theta \pm \alpha_j \sin\theta \ .
    \end{aligned}
    \right.
\end{equation}
By evaluating $\det A$ and $\text{tr}A$ for each on-meridian equilibrium it is possible to conclude that the ``up'' and ``down'' equilibria are stable foci, while the $S_{1,2}$ equilibria are unstable saddle points. The $\bar S_{M}$ curve for various spin torque strengths are shown in Fig.\ \ref{fig_astroid_1}.

When magnetic field $\bf h$ crosses $\bar S_M$, one of the stable equilibria $M_{up/down}$ merges with one of the saddles $S_{1,2}$ and disappears. In fact, $\bar S_M$ represents the Stoner-Wohlfarth (SW) astroid boundary modified by the current:\cite{PhysRevB.83.054425} The original astroid shape is squeezed along one of the diagonal directions. The segments of $\bar S_M$ connecting the corner points of the astroid become unequal: two of them grow with increasing $\alpha_j$, and the other two shrink. The details of merging depend on whether the long or the short segment is crossed by the field. On the short segment the sign of the $m_z$ component of the disappearing equilibrium is always opposite to the sign of the field component $h_z$. This property was satisfied everywhere on the conventional SW astroid boundary, and we denote the short segment of $\bar S_M$ as $\bar S_{Mc}$ with index ``c'' meaning ``conventional''. On the long segment the sign of $m_z$ is not determined by the sign of $h_z$. Indeed, points $M_{up}$ and $S_1$ merging on this segment have $m_z > 0$, and at the same time it can be crossed by a field with $h_z > 0$, if $\bf h$ is directed at a small enough angle to the $y$ axis. We denote the long segment as $\bar S_{Mu}$ with index ``u'' meaning ``unconventional''.

\subsubsection{Off-meridian equlibria}\label{sec:off_meridian}
\begin{figure}[t]
    \centering
    \includegraphics[width=0.35\textwidth]{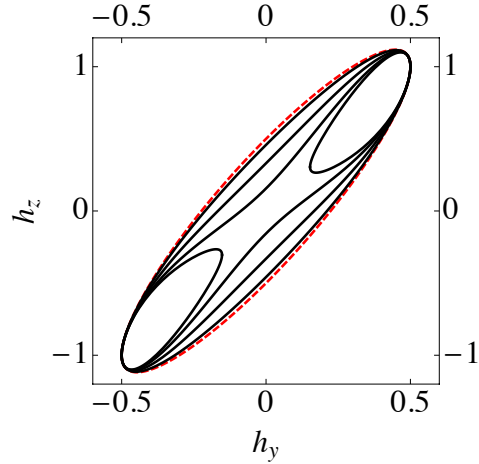}
    \caption{(color online) The $\bar S_{OM}$ boundaries for various damping factors with spin torque strength set to $\alpha_j=0.5$. The damping parameters $\alpha/\alpha_j$ are set to be 0.8, 1.5, 1.9, and 2.1 (going from the outermost to the innermost curves). At low damping regime, $\bar S_{OM}$ is approaching the tri-equilibria merging boundary which is shown as a red (grey) dashed curve in the figure.}
    \label{fig_astroid_2}
\end{figure}

Next, we analyze the stability of the off-meridian equilibria. According to Eq.\ (\ref{l_pm}) $\lambda \leqslant \lambda_{-}$ is automatically satisfied when $\lambda_\pm$ are real. Thus, according to criteria (\ref{stb}), only the  $\bar S_T$ critical line is relevant whether $\lambda_\pm$ are real or complex, and the stability condition is given by
\begin{equation}\label{ineq}
    \lambda - \lambda_{\text{T}} = \left(\frac{\alpha_j}{\alpha}\cos\phi - \frac{1}{2}\sin\theta\right)\sin\theta > 0.
\end{equation}
This inequality can be satisfied only for $\cos\phi > 0$, which means that the $OM_2$ equilibrium characterized by $\pi/2 \leq \phi \leq 3\pi/2$ is always unstable. The off-meridian equilibrium with $-\pi/2 < \phi < \pi/2$ can be stable. The critical line $\lambda = \lambda_T$ gives a destabilization boundary $\bar S_{OM}$ for this equilibrium. Its analytic form is obtained from Eqs.\ (\ref{h_y}) and (\ref{h_z}) as
\begin{equation}\label{cf3}
    h_z = \frac{h_y}{\alpha_j} \pm \sqrt{{\alpha_j}^2 - {h_y}^2} \sqrt{1 - \frac{\alpha^2}{4{\alpha_j}^2} \left(1 - \frac{{h_y}^2}{{\alpha_j}^2} \right)}
\end{equation}
The $\bar S_{OM}$ curve for various damping parameters at a fixed spin torque strength is shown in Fig.\ \ref{fig_astroid_2}. From $\det A$ and $\text{tr}A$ analysis it is possible to extract more detailed information about the nature of the $OM_{1,2}$ equilibria. We find that $OM_2$ is always an unstable node (two real positive eigenvalues), and $OM_1$ is a stable focus (complex conjugate eigenvalues with negative real parts) inside its domain of stability bounded by the $\bar S_{OM}$ line (see Appendix \ref{off_meridian_stability_analysis} for a complete analysis). As the field increases and moves out of this domain, $OM_{1}$ is destabilized but not destroyed. It continues to exist, first as an unstable focus, and then as an unstable node, until it finally merges with the points $OM_2$ and $S_2$, as discussed in Sec.~\ref{sec:EvolutionEquilibriumStates}. More details of the $OM_1$ state evolution are provided in Appendix~\ref{off_meridian_stability_analysis}.

\subsection{Current-field diagrams}

\begin{figure}[t]
    \centering
    \includegraphics[width=0.3\textwidth]{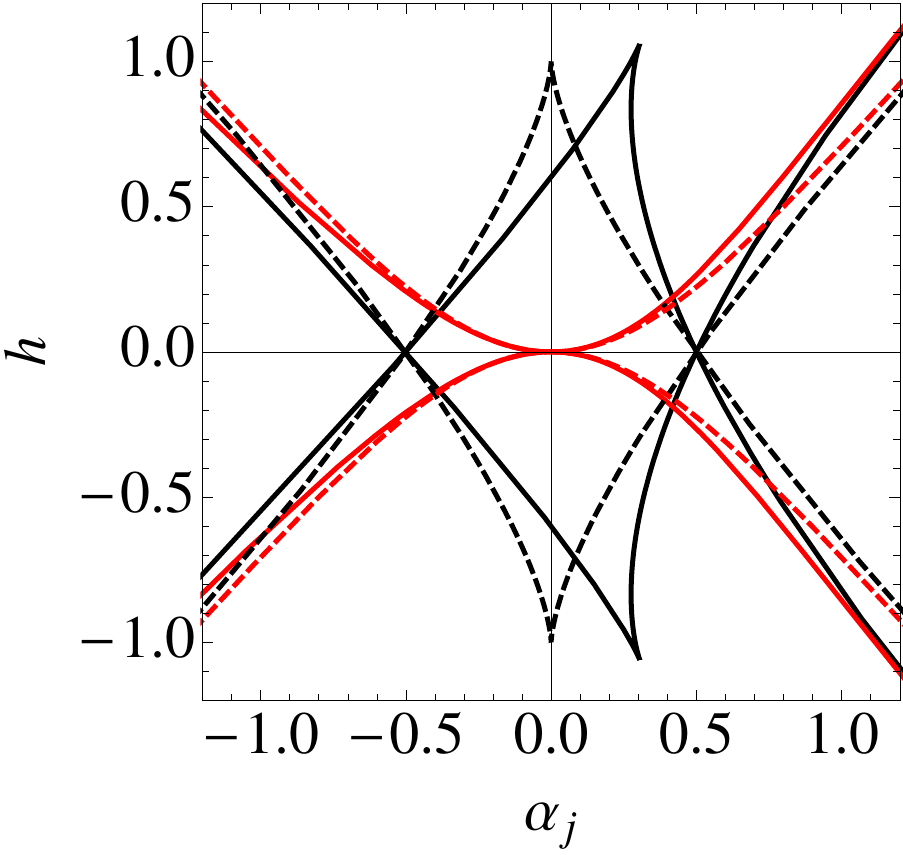}
    \caption{(color online) Tilted current-field phase diagram at $\alpha = 0$. The black and red(gray) lines correspond to $h_{cM}$ and $h_{cOM}$, respectively. The solid lines represent the diagrams with a tilting angle of the field set to be 15 degree with respect to $\hat y$ direction. The dashed lines represent the diagram with ${\bf h} || \hat y$. Opposite tilting happens when the angle is negative.}
    \label{fig_current_field}
\end{figure}

Switching phase diagrams Fig.\ \ref{fig_astroid_1} and Fig.\ \ref{fig_astroid_2} describe experiments performed at fixed current with magnetic field of a fixed direction increased until switching happens at a critical value $h_c$. In a more complicated experiment one can measure how the $h_c$ threshold depends on the current magnitude. Such experiments were indeed recently performed by Yu {\em et al}.\cite{PhysRevB.89.104421} and investigated numerically.\cite{chang2011multiple} The $h_c(\alpha_j)$ dependencies were measured for different field directions and, quite surprisingly, it was found that for fields making a finite angle with the $y$ axis multiple switchings may occur. This fact finds a natural explanation in the framework of our theory.  In terms of Fig.~\ref{fig_astroid_1}
the critical fields are determined by intersections of a straight line representing fields of given direction with the lines $\bar S_M$ and $\bar S_{OM}$. If the direction of the field is defined by the angle $\theta_h$ with the $z$-axis, the former intersection point $h_{cM}(\alpha_j)$ can be found by solving Eqs.~(\ref{astroid}) with $h_y = h_{c} \sin\theta_h$, $h_z = h_{c} \cos\theta_h$. For the latter intersection point $h_{cOM}(\alpha_j)$ Eqs.~(\ref{cf3}) should be used. The results are shown in Fig.\ \ref{fig_current_field}. One can see that $h_{cM}(\alpha_j)$ exhibits a sharp peak located at $\alpha_j = 0$ for $\theta_h = \pi/2$. As the field is tilted away from the $y$-axis, the position of the peak moves and its initially symmetric shape deforms. Eventually the deformation grows so big that the function $h_{cM}(\alpha_j)$ becomes multi-valued, in accord with experimental findings. Comparing the current-field diagram with the experimental diagram (Fig.\ 3 of Ref.~\onlinecite{PhysRevB.89.104421}) one can see a good qualitative correspondence.

Here we show how the shape of $h_{cM}(\alpha_j)$ can be understood from the evolution of the modified astroid $\bar S_M$.  As the current is increased with $\alpha_j > 0$, the astroid is squeezed in the $(1,-1)$ direction and expanded in the $(1,1)$ direction. The $\bar S_{Mu}$ lines approach the origin in the $h$-space and the $\bar S_{Mc}$ lines move away from it. When the field is directed along the $y$ axis, its line intersects  the $\bar S_{Mu}$ boundary. Since this boundary moves towards the origin with increasing $\alpha_j$, the function $h_{cM}(\alpha_j)$ is decreasing. However, when the field is directed at an angle to the $y$-axis, its line may initially cross the $\bar S_{Mc}$ boundary. Since $\bar S_{Mc}$ moves away from the origin, the function $h_{cM}(\alpha_j)$ would increase. At a threshold value of current the field line goes exactly through the corner point between the $\bar S_{Mc}$ and $\bar S_{Mu}$. At this point $h_{cM}(\alpha_j)$ exhibits a cusp. For currents above the threshold, the field line crosses $\bar S_{Mu}$ and, just like in the ${\bf h} || \hat y$ case, $h_{cM}(\alpha_j)$ becomes a decreasing function. For some angles $\theta_h$ there may be situations when the field line crosses both $\bar S_{Mc}$ and $\bar S_{Mu}$ lines. This is when $h_{cM}(\alpha_j)$ becomes multi-valued and complicated hysteresis patterns are realized.

The form of the other critical field line, $h_{cOM}(\alpha_j)$  (red curves in Fig.~\ref{fig_current_field}), is related to the evolution of the $\bar S_{OM}$ line. Since this line moves away from the origin in all directions, $h_{cOM}(\alpha_j)$ turns out to be an increasing function.

\subsection{Discussion of the phase diagram}

The $\bar S_M$ and $\bar S_{OM}$ lines together give the complete switching phase diagram in the $y-z$ field plane. For small values of $\alpha_j $ the critical line $\bar S_M$ is qualitatively equivalent to the conventional Stoner-Wohlfarth astroid, and the equilibrium merging process is similar: There are four on-meridian equilibria for $\bf h$ inside the astroid, and as the field crosses its boundary two of them merge and disappear. Above the critical current $\alpha_j = 1/2$, the $\bar S_M$ critical line becomes self-crossing (Fig.\ \ref{fig_astroid_1}). At the critical current the $\bar S_{Mu}$ lines touch each other at ${\bf h} = 0$ so the threshold can be found from Eq.~(\ref{astroid}) with $h_y = h_z = 0$.  Inside the region of self-crossing there are no on-meridian equilibria, as already observed in Ref.~\onlinecite{PhysRevB.83.054425} and the supplemental material of Ref.~\onlinecite{PhysRevLett.109.096602}. However, the Poincar\'{e}-Hopf theorem is not violated due to the presence of the off-meridian equilibria.

In the absence of current the system in constant external field $\bf h$ resides in one of the two stable $M$ equilibria. As $\alpha_j$ is increased, the oval-shaped region of stability of the $OM_1$ state grows, and the area inside the modified astroid $\bar S_M$ shrinks. Moreover, the self-crossing region of $\bar S_M$, where no on-meridian equilibria exist, also grows. Thus both $M_{up}$ and $M_{down}$ states eventually become unstable at some critical current $\alpha_j^M$ and $\bf m$ switches to the $OM_1$ state. What happens if the current is subsequently decreased? The answer to this question can be read from the $h_c(\alpha_j)$ dependence shown in Fig.\ \ref{fig_current_field}. At a given $h$ the off-meridian state remains stable down to the current $\alpha_j^{OM}$ obtained from the equation $h = h_{cOM}(\alpha_j)$. If $\alpha_j^{OM} < \alpha_j^{M}$, one would observe hysteretic behavior of the system in the current interval between $\alpha_j^{OM}$ and $\alpha_j^{M}$. At the higher end of this interval the system switches from an $M$ state to the $OM_1$ state. At the lower end it switches back to an $M$ state. As seen from Fig.\ \ref{fig_current_field}, the length of the bistable interval becomes larger for smaller $h$. At ${\bf h}=0$  one finds using Eq.~(\ref{cf3}) that $OM_1$ is stable for $\alpha_j > \alpha/2$. The higher end of the interval was already discussed --- it corresponds to the first self-crossing of $\bar S_M$, i.e., to $\alpha_j = 1/2$. For typical values of Gilbert damping $\alpha \sim 0.01$ the hysteresis range is very large. It requires an initial pulse of current of the order $\alpha_j \sim 1$ to get to the $OM$ state, but after that the current can be reduced to $\alpha_j \sim \alpha$, and the $OM$ state can be comfortably studied at low currents. Experiments with SHE devices\cite{Miron2011,PhysRevLett.109.096602} are already performed in the regime $\alpha_j \sim 1$ so the discussed hysteresis should be observable.

When magnetic field is set inside the domain of existence of $OM$ states but outside of their domain of stability, the system has two unstable $OM$ equilibria. It is possible to arrange parameters so that there no $M$ equilibria either (this happens in the high damping, high current regime). In this case the system has no choice but to follow some precession cycle, the analysis of which is beyond the scope of the present paper.

\section{Dynamic properties}

In this section we discuss what happens after the stability boundaries are crossed and equilibria are destabilized.

\subsection{Switching to the off-meridian state}
\begin{figure*}
    \centering
    \subfigure{\label{fig_phase_a}\includegraphics[width=0.3\textwidth]{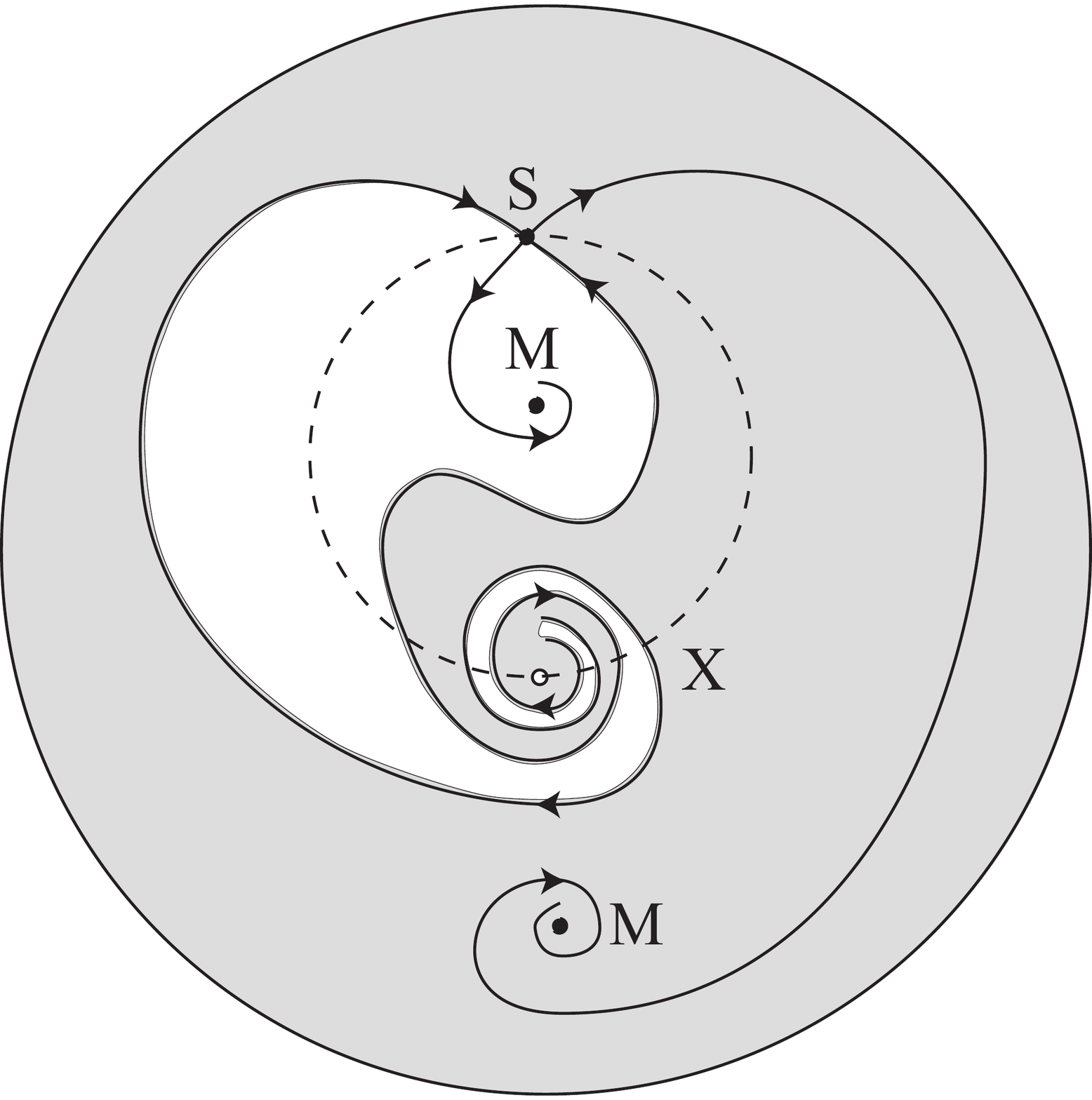}}\hfill
    \subfigure{\label{fig_phase_b}\includegraphics[width=0.3\textwidth]{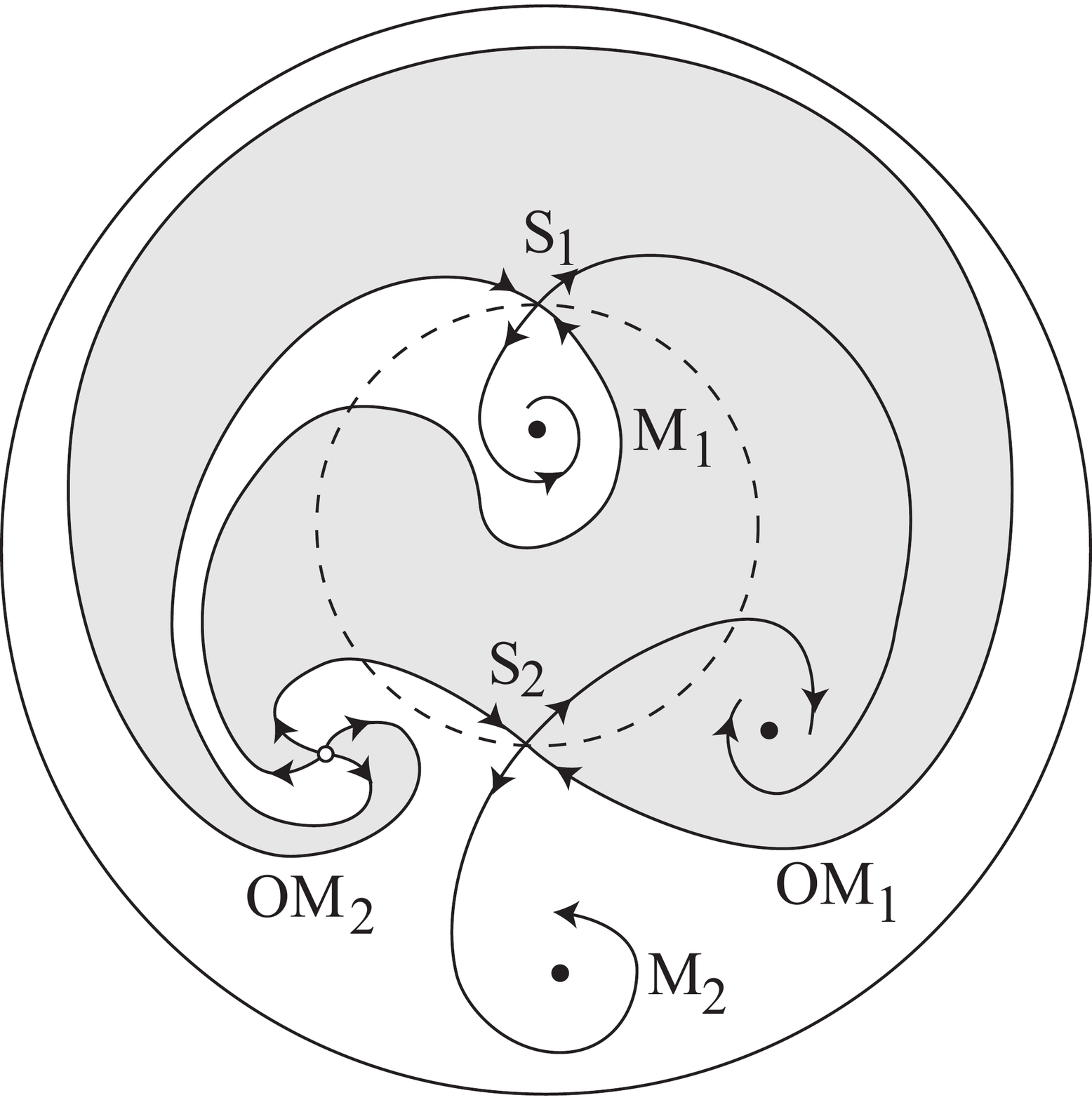}}\hfill
    \subfigure{\label{fig_phase_c}\includegraphics[width=0.3\textwidth]{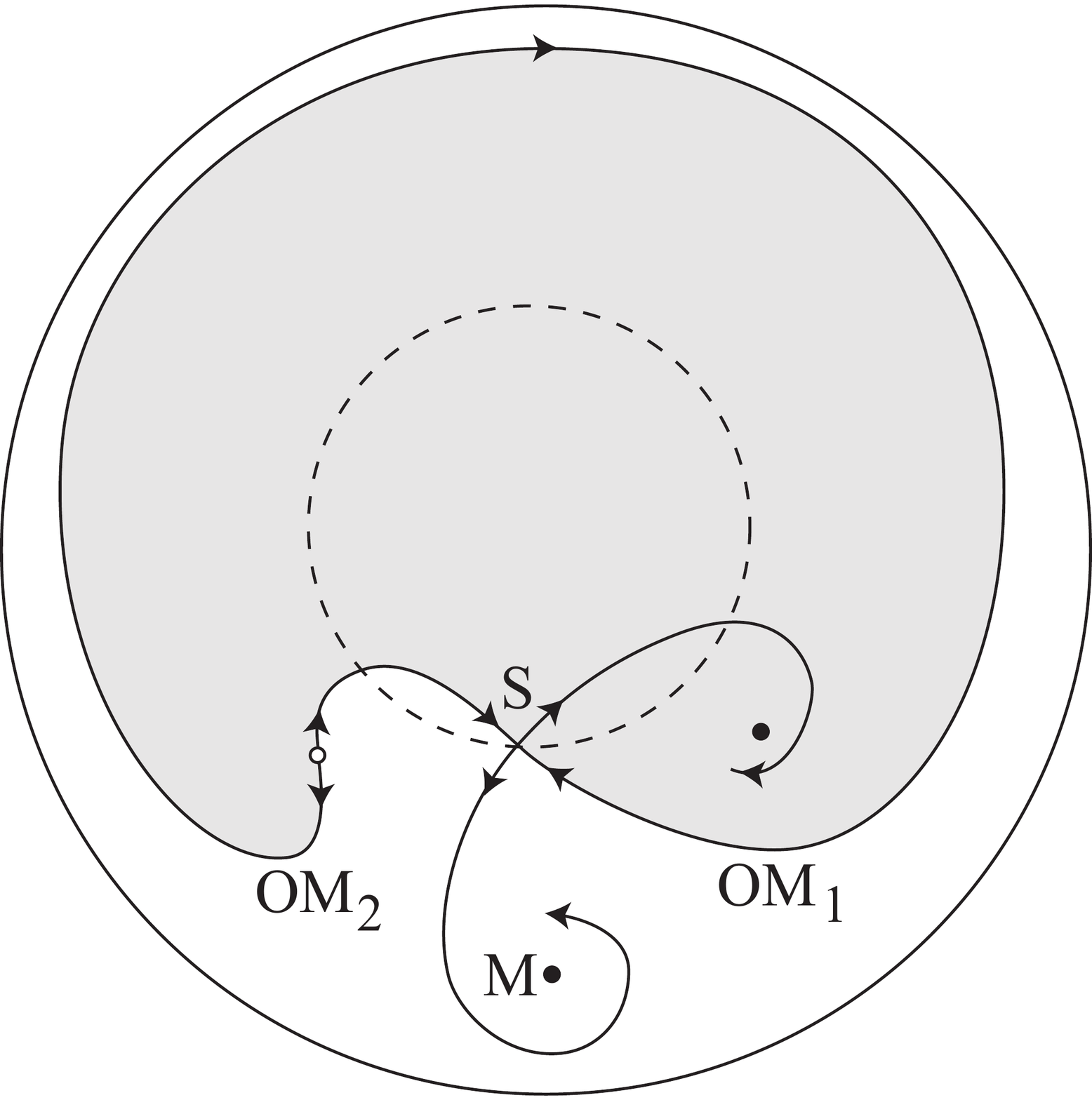}}
    \caption{(color online) Sketches of the phase portraits on the stereographic plane. The dashed circle in each subfigure is the projection of the equator. The north and south poles are projected to the circle center and to infinity, respectively}\label{fig_phase}
\end{figure*}

Existence of a stable $OM$ state within the area given by Eqs.~(\ref{cf3}) raises a question: When an $M$ state is destabilized at the $\bar S_M$ boundary, will the system switch to the other $M$ state, or to a stable $OM$ state? To answer this question we plot the flow diagrams (phase portraits) of the system. The results of calculations are presented in the form of qualitative sketches that emphasize the structure of the flow (Fig.\ \ref{fig_phase}). In the field-dominated regime the flow is qualitatively similar to that in the absence of the current. There are two basins of attraction of stable points $M_{up}$ and $M_{down}$ (white and dark areas in Fig.\ \ref{fig_phase}a). The separatrix between the two basins winds around the unstable focus $X$ making infinite number of turns. As a result, near $X$ the basins are finely intermixed and a small change in initial conditions may change the equilibrium where the system ends up. When modified astroid boundary is crossed, one of $M$ points is destroyed. A system initially residing in this point will switch to the other $M$ point.

In the current-dominated regime there are three basins of attraction. The one of the $OM_1$ point (dark area in Fig.\ \ref{fig_phase}b) separates those of $M_{up}$ and $M_{down}$ (white areas). The white areas touch at the point of unstable equilibrium $OM_2$. The important difference from the field-dominated regime is that $OM_2$ is an unstable node, and not a focus. Thus, there is no winding of the separating line around it and, consequently, no area of fine intermixing. Fig.\ \ref{fig_phase}c shows what happens when $M_{up}$ and $S_1$ points merge at the modified astroid boundary. The phase portrait in the upper part of the unit sphere qualitatively changes: The basins of attraction of $M_{up}$ and $OM_1$ merge, forming one bigger basin of attraction of $OM_1$. This transformation of the phase portrait does not affect the qualitative picture in the lower part of the unit sphere and the boundary between the basin of $OM_1$ and $M_{down}$. The end result is that a system initially residing in $M_{up}$ will switch to the $OM_1$ state with certainty. The latter statement, of course, only applies to the case of slow, quasi-static change of parameters, in which case one can be sure that $\bf m$ follows the stable point with great accuracy. If parameters are changed at a finite speed, there will be a lag between $\bf m$ and the equilibrium point, and a more careful investigation would be required.

\subsection{Two-stage switching through the off-meridian equilibrium}
Magnetization reversal is one of the most important processes in magnetism that is linked to the magnetic data storage process, such as in hard disk drives. Switching speed and reliability are two crucial factors to the design of such systems. In conventional spin-transfer torque switching spin polarizer is directed along the easy axis of the free layer. Then the magnetization moves towards the new equilibrium along a spiral trajectory in a reliable but fairly slow manner.\cite{bertotti2003comparison, ralph2008spin, bazaliy2011ballistic, bazaliy2011analytic, koch2004time, sun2000spin} Much faster reversals, which are often called precessional switchings, have been designed. Some have magnetic field applied orthogonally to the easy axis. Others use spin polarizer perpendicular to the easy plane of the free layer (``magnetic fan'' geometry).\cite{bauer2000switching,kent2004spin,liu2010ultrafast} In both cases the reversal processes is fast but requires current or field pulse length to be perfectly adjusted. This is experimentally hard to control and increases the error rate.

\begin{figure}
\centering
\includegraphics[scale=0.8]{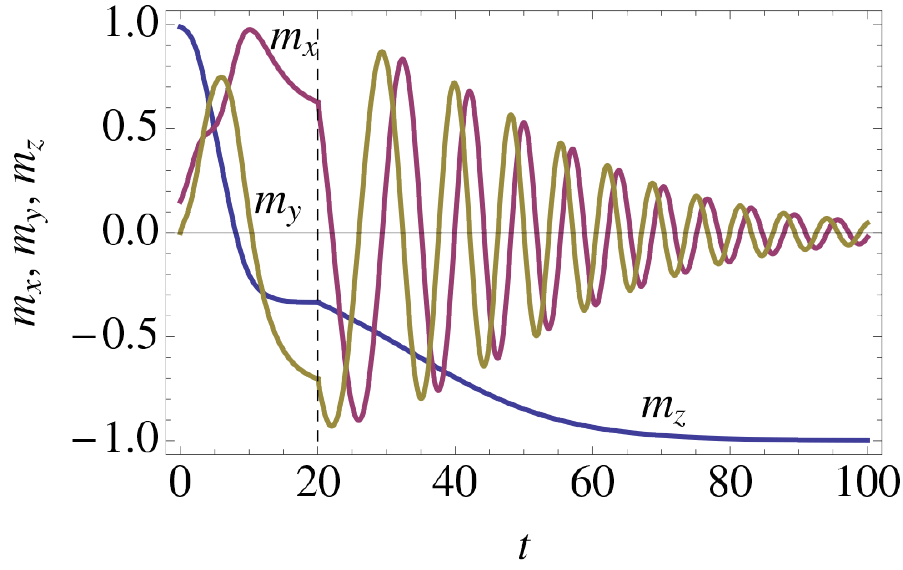}
\caption{(color online) Evolution of the three components of the magnetization in a two stage switching process. The parameters are: \(h_y=0.15\), \(\alpha_j=0.45\), and \(\alpha=0.05\). The field-current pulse is turned on at \(t=0\) and off at \(t=20\).}
\label{fig:oop_switch}
\end{figure}

It was numerically found in Refs.~\onlinecite{lee2013threshold} and \onlinecite{chang2011multiple} that switching from an $M$ state to the $OM$ state is fast and does not exhibit precession. A recent micromagnetic simulation paper has reported the similar switching behaviors, which indicates the validity of the OOP switching in the microspin regime.\cite{finocchio2013switching}. Figure~\ref{fig:oop_switch} shows the process of switching from $M_{up}$ to $OM_1$ state. It is seen that the switching time is of the order of ferromagnetic resonance period $T$ ($T = 2\pi$ in dimensionless unit used in Fig.~\ref{fig:oop_switch}). We will consider one possible scenario of switching from $M_{up}$ to $M_{down}$ equilibrium with an intermediate stop in the $OM_1$ state. Consider a system that is initially in the $M_{up}$ state. Magnetic field $\bf h$ is set in the negative $z$ direction during the whole switching procedure with $h < 1$ so that $M_{up} = +\hat z$ is stable. First, we apply a short pulse of strong current $\alpha_j \sim 1$. The rise and fall times of the pulse are assumed to be negligible. During the pulse time $M_{up}$ does not exist and $\bf m$ switches to $OM_1$. The pulse length $t_p$ is selected to be large enough for the switching process bo be accomplished. Importantly, this requirement sets only a low bound for $t_p$ --- there will be no harm in keeping the current on for a longer time. According to Eqs.~(\ref{cf3}) for $h_y = 0$ the state $OM_1$ has $\theta = \pi/2$ and sits on the equator of the unit sphere. After the end of the pulse the current is switched off and the second stage of switching begins. Now the states $M_{up}$ to $M_{down}$ are stable again and $\bf m$ should go to one of them. With field pointing down and $\alpha_j = 0$, the boundary between the basins of attraction of $M_{up}$ and $M_{down}$ is a parallel circle, located above the equator of the unit sphere. Thus the second stage starts with $\bf m$ residing in the basin of attraction of $M_{down}$, to which $\bf m$ eventually relaxes in a precessional manner. The whole process is characterized by a fast first stage with strong current and a slow second stage, during which the systems is not driven externally. While the total switching time is of the same order of magnitude as in the conventional switching, the ``active'' stage requires much shorter time, comparable to that of precessional switching, making the procedure potentially useful for special applications. An important drawback of this switching scenario is that for a given direction of $\bf h$ it can be performed only in one direction, e.g., in the discussion above from $M_{up}$ to $M_{down}$. To switch back one would have to reverse the direction of $\bf h$.

It is interesting to to note that the SHE device switching between $M_{up}$ and $M_{down}$ in a two-stage manner described above can be alternatively viewed as a realization of a controlled-NOT gate with $h_z$ being the control parameter.

Finally, we want to compare the duration of the fast stage of SHE switching with the switching time of a conventional spin-torque device, where the magnetic polarizer and the external field are both pointing along the easy axis of the free layer. Assuming the conventional spin torque to be described by a constant spin-transfer efficiency factor $g(\theta) = \bar g$, we get ${\bf h}_{sp} = \alpha_j [{\bf m} \times \hat z]$ with $\alpha_j=\bar{g}j/j_0$ for its effective field. In this fully axially-symmetric case the switching time can be computed analytically\cite{yan2014nonlinear} as
\begin{equation}\label{eq:conentional_ts}
    t_{s} = \frac{1}{2\alpha (1-h')}\ln\left( \frac{1-m_{z0}}{1- m_{z0}/h'}  \right),
\end{equation}
with $h' = h + \alpha_j/\alpha$ and $m_{z0}$ being the initial value of the magnetization component along the easy axis. In the small-damping ($\alpha \ll 1$), large-current ($\alpha_j/\alpha \gg h \sim 1)$, regime this simplifies to $t_s \approx -\ln(1-m_{z0})/(2\alpha_j)$. Conventional switching requires some initial deviation of $\bf m$ from equilibrium. This deviation is usually thought to come from thermal fluctuations and can be evaluated by using Maxwell equilibrium distribution for  $m_{z0}$
\begin{equation}\label{eq:dev}
    \rho(m_{z0})=\kappa\sqrt{1-m_{z0}^2}e^{-E(m)/{k_bT}}.
\end{equation}
where $\kappa$ is the normalizing constant. The typical expected value at room temperature is around $m_{z0} \approx 0.99$ ($\theta_0 \approx 0.5^{\circ}$).\cite{d2004nonlinear} To compare the switching times it is important to remember that conventional and SHE devices differ in two aspects. On the theoretical level, in conventional devices switching occurs at $\alpha_j \sim \alpha$, while $\alpha_j \sim 1$ is required for SHE switching. On a practical level conventional devices can bear smaller currents due to heating problems. Thus achieving $\alpha_j \sim 1$ in them is problematic. In view of that, we perform two comparisons. First, we compare the SHE and conventional switching times for $\alpha_j = 0.5$ and  $h=0$. Here we get $t_p \approx 14$ and $t_s \approx 5$. Given the same normalized spin torque strength, conventional device is faster than the SHE one. Second, we compare the two devices operating at their critical switching current with a small field, say $h=0.02$, pointing toward the $-z$ direction. For the SHE device we again use $\alpha_j = 0.5$ and the resulting
switching time does not change much, $t_p \approx 13.5$. For a conventional devices we use $\alpha_j = \alpha$, then $t_s \approx 29/\alpha$. In this sense the SHE switching turns out to be much faster. In addition, since the initial condition is a statistical average, the switching time estimated in this fashion may cause an non-negligible error in experiments.

\section{Summary}

The method of Refs.~\onlinecite{PhysRevB.61.12221, PhysRevB.88.054408} provides a framework that can be applied to find the critical switching surfaces for any magnetic system with arbitrary anisotropy and spin torque in an exact fashion. In this article we calculated the three-dimensional critical surface for a SHE bilayer system with perpendicular anisotropy and in-plane current. For external fields in the $x-y$ plane, the SHE induced spin torque not only shifts the existing equilibria, but also generates two new off-meridian equilibria. 

{\em First}, we derive an analytic formula for the stability boundary for the off-meridian equilibria --- in previous numeric research this boundary was not distinguished from the boundary their existence. {\em Second}, in contrast to the other authors we discuss the switching phase diagrams in the field space at a constant current, and plot the modified astroid and the oval-shaped stability region of the off-meridian state. We then show how our qualitative description of the evolution of the constant current switching boundaries can explain the results of the other authors obtained for variable currents. {\em Third}, we discuss in detail the evolution of equilibrium points and the character of their destabilization on the switching boundaries. This allows us to put forward a qualitative understanding of the complicated hysteresis processes that are found in SHE devices. {\em Fourth}, we point out that while a large current is required to set magnetization into the off-meridian state, it remains in this state when current is decreased to values that are $\alpha$ times smaller, and thus can be studied at low currents. {\em Fifth}, we show that in the presence of stable off-meridian state (current-dominated regime) switching between ``up'' and ``down'' equilibria happens through the off-meridian state. We consider an example of such two-stage switching and find the relationship of the corresponding switching time with that in conventional spin-torque device with magnetic polarizer. Here we find that, depending on the current limitations not related to spin torque physics, either of the two devices can operate faster.

\begin{acknowledgments}
    This research was supported by National Science Foundation Grant No. DMR-0847159.
\end{acknowledgments}

\appendix

\section{}\label{mtxelms}
The explicit expression of each component in Eq.\ (\ref{vecfld}) can be derived as
\begin{subequations}
    \begin{align}
        \partial_\theta f^\theta =&-\partial _{\theta \theta }\varepsilon +\partial _{\theta }h_{\text{sp}}{}^{\theta },\\
        \partial_\theta f^\phi =&-\partial _{\theta }\left(\frac{1}{\sin  \theta }\partial _{\phi }\varepsilon \right)+\partial _{\theta }h_{\text{sp}}{}^{\phi},\\
        \partial_\phi f^\theta =&\frac{\cos  \theta }{\sin  \theta }\left(\frac{1}{\sin  \theta }\partial _{\phi }\varepsilon -h_{\text{sp}}{}^{\phi }\right)
        \nonumber
        \\
        &\quad -\frac{1}{\sin \theta }\left(\partial _{\theta \phi }^2\varepsilon - \partial _{\phi }h_{\text{sp}}{}^{\theta }\right),\\
        \partial_\phi f^\phi =&-\frac{\cos  \theta }{\sin  \theta }\left(\partial _{\theta }\varepsilon -h_{\text{sp}}{}^{\theta }\right)
        \nonumber
        \\
        &\quad -\frac{1}{\sin ^2\theta}\partial _{\phi \phi }^2\varepsilon +\frac{1}{\sin  \theta }\partial _{\phi }h_{\text{sp}}{}^{\phi}.
    \end{align}
\end{subequations}

\section{}\label{table}

For a planar linear system of the form $\dot{\bf{X}} = A\bf{X}$, the eigenvalues of the $2 \times 2$ coefficient matrix $A$ can be calculated in terms of its trace and determinant as\cite{hirsch2004differential}
\begin{equation}
    \mu_\pm = \frac{1}{2}\left(\text{tr}A \pm \sqrt{(\text{tr}A)^2-4\text{det}A}\right).
    \label{eq:eigenvalue}
\end{equation}
Therefore knowing $\text{tr}A$ and $\text{det}A$ tells us virtually everything about the geometry of its solutions.

Besides stability, the types of an equilibrium is also of importance in understanding the switching process. An equilibrium of the same stability (except saddle point) can be a node or a focus, depending on whether the eigenvalues (\ref{eq:eigenvalue}) are real or complex. Therefore the differentiation of the focus set and the node set requires another pair of critical values $\lambda_{c\pm}$ which satisfies
\begin{equation}
    (\text{tr}A)^2-4\text{det}A = 0.
    \label{eq:node_focus}
\end{equation}
It can be demonstrated that the relationships $\lambda_{c+} \geqslant \lambda_+$ and $\lambda_{c-} \leqslant \lambda_-$ are always satisfied.

The classification of stability and equilibrium type in terms of the eigenvalues and $\lambda$ are summarized in TABLE \ref{table:eq} and in TABLE \ref{table:type}, respectively. The complete dynamic analysis of an equilibrium needs to take into account the two factors together.

\begingroup
    \squeezetable
    \begin{table*}
        \caption{Classification of stability}\label{table:eq}
        \centering
        \begin{tabular}[b]{l | l | l | l | l}
            \hline\hline
            Stability Type & Equilibrium Type & Eigenvalue Equivalent& $\text{tr}A$-$\text{det}A$ Equivalent & $\lambda$ Equivalent \\
            \hline
            \multirow{2}{*}{Sink} & Stable focus & Complex $\text{Re}[\mu_\pm ] < 0 $ &
            \multirow{2}{*}{$\text{tr}A < 0$, $\text{det}A > 0$} &
            \multirow{2}{*}{$\left\{
             \begin{array}{l l}
                \lambda > \text{Max}(\lambda_\text{T},\lambda_{+}) & \text{if } \lambda_\text{T} \geqslant \lambda_{-}\\
                \lambda_\text{T} < \lambda < \lambda_{-} & \text{if } \lambda_\text{T} < \lambda_{-}
             \end{array} \right.$}  \\
            & Stable node & Real $\mu_- < \mu_+ < 0$ &  &
            \\ \hline
            \multirow{2}{*}{Source} & Unstable focus & Complex $\text{Re}[\mu_\pm ] > 0 $ &
            \multirow{2}{*}{$\text{tr}A > 0$, $\text{det}A > 0$} &
            \multirow{2}{*}{$\left\{
             \begin{array}{l l}
                \lambda < \text{Min}(\lambda_\text{T},\lambda_{-}) & \text{if } \lambda_\text{T} \leqslant \lambda_{+}\\
                \lambda_\text{+} < \lambda < \lambda_{T} & \text{if } \lambda_\text{T} > \lambda_{+}
             \end{array} \right.$} \\
            & Unstable node & Real $0 < \mu_- < \mu_+$ &  &
            \\ \hline
            Saddle & Saddle point & Real $\mu_- < 0 < \mu_+$ & $\text{det}A < 0$ & $\lambda_- < \lambda < \lambda_+ $ \\
            \hline\hline
        \end{tabular}
    \end{table*}
    \begin{table*}
        \caption{Classification of focus and node}\label{table:type}
        \centering
        \begin{tabular}[b]{l | l | l | l}
            \hline\hline
            Equilibrium Type & Eigenvalues & $\text{tr}A$-$\text{det}A$ Equivalent & $\lambda$ Equivalent \\
            \hline
            Node & Real & $(\text{tr}A)^2-4\text{det}A > 0$, $\text{det}A > 0$ & $\lambda_{c-} < \lambda < \lambda_-$ or $\lambda_+ < \lambda < \lambda_{c+}$ \\
            \hline
            Focus & Complex & $(\text{tr}A)^2-4\text{det}A < 0$ \footnote{Note that $(\text{tr}A)^2-4\text{det}A < 0$ guarantees $\text{det}A > 0$.} & $\lambda < \lambda_{c-}$ or $\lambda > \lambda_{c+}$ \\
            \hline\hline
        \end{tabular}
    \end{table*}
\endgroup

\section{}\label{off_meridian_stability_analysis}

We mentioned the equilibrium types of the two off-meridian equilibrium states and their evolution in Sec.~\ref{sec:EvolutionEquilibriumStates} and in Sec.~\ref{sec:off_meridian}. To quantitatively understand the evolution of these two states, we also need to find the critical $\lambda$ which separates nodes and foci, i.e., to solve Eq.~(\ref{eq:node_focus}). It's solution in the off-meridian case is given by
\begin{eqnarray}
    \lambda_{\text{c}\pm} =&& \frac{\sin ^2 \theta}{2} - \cos ^2 \theta \pm \frac{1}{2}\sin ^2 \theta \sqrt{1 + \alpha ^2}
    \nonumber
    \\
    && \qquad + \alpha_j \alpha \sin \theta \cos \phi.
\end{eqnarray}

The two critical values $\lambda_{\text{c}\pm}$ give another two surfaces in the parameter space, one above $\lambda_+$ and the other below $\lambda_-$. Since both the two off-meridian equilibria have $\lambda \equiv -\cos ^2 \theta < \lambda_-$, we only need $\lambda_{\text{c}-}$ to determine the equilibrium type.

\begin{figure}[h]
    \centering
    \includegraphics[width=0.35\textwidth]{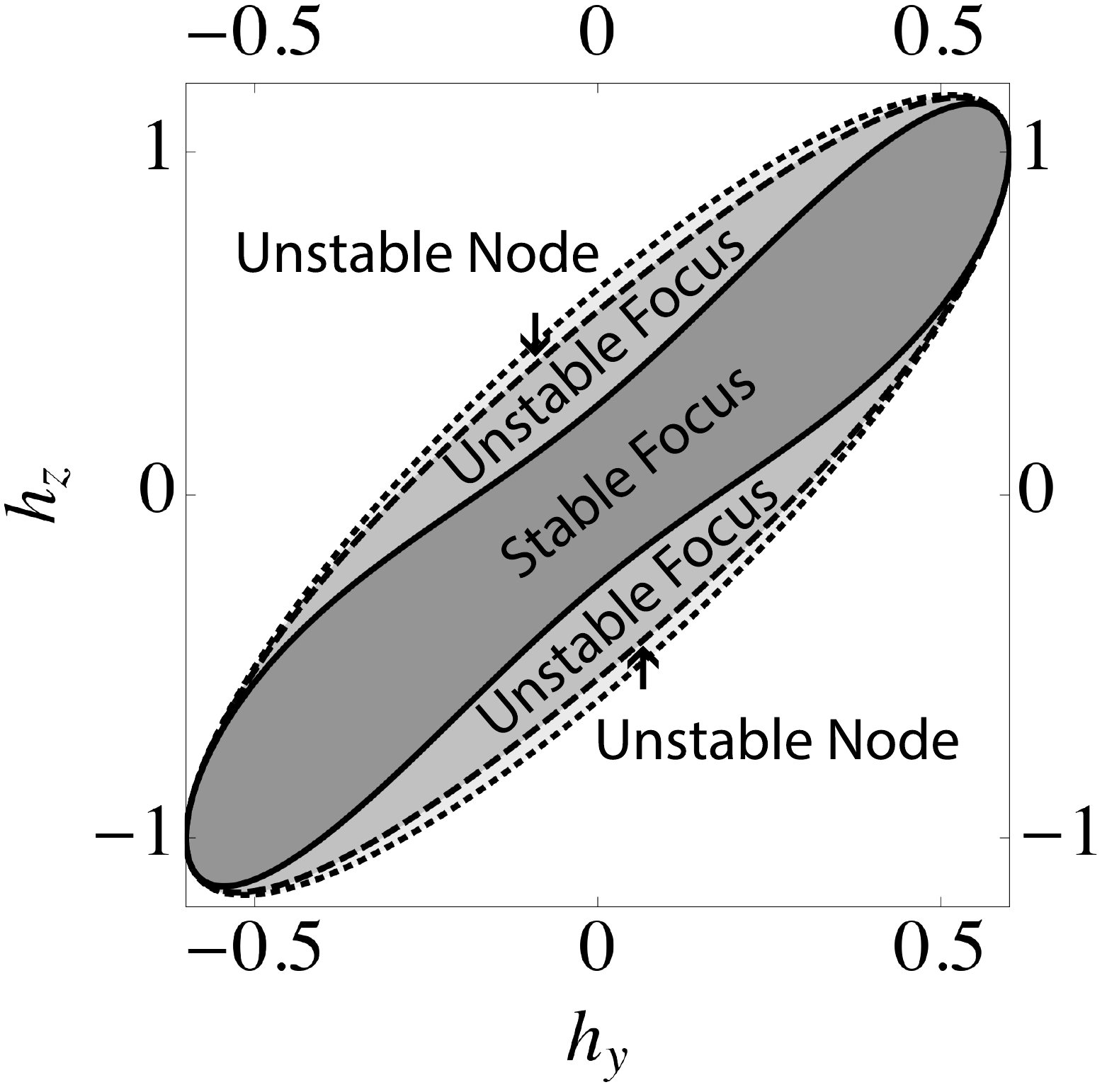}
    \caption{The $\bar S_{OM}$, $\bar S_{c}$, and the equilibrium merging boundary are plotted with solid, dashed, and dotted lines, respectively. The unstable node, unstable focus, and stable focus regions are marked in different grayscales. For illustrative purposes, we adopt $\alpha_j= 0.6$ and $\alpha = 1$ to make the unstable regions large enough to be seen.}
    \label{fig:node_focus_region}
\end{figure}

The two off-meridian equilibrium have different equilibrium types. For the one with $\pi/2 < \phi < 3\pi/2$, we have $\lambda_- \geqslant \lambda \geqslant \lambda_{\text{c}-} $, therefore it is always an unstable node. The other one which satisfies $-\pi/2 < \phi < \pi/2$ may change the equilibrium type as field changes. we can find two critical curves by equating $\lambda$ to $\lambda_-$ and to $\lambda_{\text{c}-}$, respectively. The former gives the destabilization boundary $\bar S_{OM}$, the latter corresponds to the type transition boundary $\bar S_{c}$ of the analytic form
\begin{equation}\label{cf4}
    h_z=\frac{h_y}{\alpha _j}\pm \sqrt{\alpha _j^2-h_y^2} \sqrt{1-\frac{\left(\sqrt{\alpha ^2+1}-1\right)^2 }{4 \alpha ^2 \alpha _j^2}\left(1-\frac{h_y^2}{\alpha _j^2}\right)}.
\end{equation}
The transition boundary $\bar S_{c}$ touches $\bar S_{OM}$ but never crosses it. It separates the unstable region of the off-meridian equilibrium into node and focus regions, as shown in Fig.~\ref{fig:node_focus_region}.

\bibliography{SHE_ref}

\begin{thebibliography}{46}%
\makeatletter
\providecommand \@ifxundefined [1]{%
 \@ifx{#1\undefined}
}%
\providecommand \@ifnum [1]{%
 \ifnum #1\expandafter \@firstoftwo
 \else \expandafter \@secondoftwo
 \fi
}%
\providecommand \@ifx [1]{%
 \ifx #1\expandafter \@firstoftwo
 \else \expandafter \@secondoftwo
 \fi
}%
\providecommand \natexlab [1]{#1}%
\providecommand \enquote  [1]{``#1''}%
\providecommand \bibnamefont  [1]{#1}%
\providecommand \bibfnamefont [1]{#1}%
\providecommand \citenamefont [1]{#1}%
\providecommand \href@noop [0]{\@secondoftwo}%
\providecommand \href [0]{\begingroup \@sanitize@url \@href}%
\providecommand \@href[1]{\@@startlink{#1}\@@href}%
\providecommand \@@href[1]{\endgroup#1\@@endlink}%
\providecommand \@sanitize@url [0]{\catcode `\\12\catcode `\$12\catcode
  `\&12\catcode `\#12\catcode `\^12\catcode `\_12\catcode `\%12\relax}%
\providecommand \@@startlink[1]{}%
\providecommand \@@endlink[0]{}%
\providecommand \url  [0]{\begingroup\@sanitize@url \@url }%
\providecommand \@url [1]{\endgroup\@href {#1}{\urlprefix }}%
\providecommand \urlprefix  [0]{URL }%
\providecommand \Eprint [0]{\href }%
\providecommand \doibase [0]{http://dx.doi.org/}%
\providecommand \selectlanguage [0]{\@gobble}%
\providecommand \bibinfo  [0]{\@secondoftwo}%
\providecommand \bibfield  [0]{\@secondoftwo}%
\providecommand \translation [1]{[#1]}%
\providecommand \BibitemOpen [0]{}%
\providecommand \bibitemStop [0]{}%
\providecommand \bibitemNoStop [0]{.\EOS\space}%
\providecommand \EOS [0]{\spacefactor3000\relax}%
\providecommand \BibitemShut  [1]{\csname bibitem#1\endcsname}%
\let\auto@bib@innerbib\@empty
\bibitem [{\citenamefont {Ando}\ \emph {et~al.}(2008)\citenamefont {Ando},
  \citenamefont {Takahashi}, \citenamefont {Harii}, \citenamefont {Sasage},
  \citenamefont {Ieda}, \citenamefont {Maekawa},\ and\ \citenamefont
  {Saitoh}}]{PhysRevLett.101.036601}%
  \BibitemOpen
  \bibfield  {author} {\bibinfo {author} {\bibfnamefont {K.}~\bibnamefont
  {Ando}}, \bibinfo {author} {\bibfnamefont {S.}~\bibnamefont {Takahashi}},
  \bibinfo {author} {\bibfnamefont {K.}~\bibnamefont {Harii}}, \bibinfo
  {author} {\bibfnamefont {K.}~\bibnamefont {Sasage}}, \bibinfo {author}
  {\bibfnamefont {J.}~\bibnamefont {Ieda}}, \bibinfo {author} {\bibfnamefont
  {S.}~\bibnamefont {Maekawa}}, \ and\ \bibinfo {author} {\bibfnamefont
  {E.}~\bibnamefont {Saitoh}},\ }\href {\doibase
  10.1103/PhysRevLett.101.036601} {\bibfield  {journal} {\bibinfo  {journal}
  {Phys. Rev. Lett.}\ }\textbf {\bibinfo {volume} {101}},\ \bibinfo {pages}
  {036601} (\bibinfo {year} {2008})}\BibitemShut {NoStop}%
\bibitem [{\citenamefont {Miron}\ \emph {et~al.}(2010)\citenamefont {Miron},
  \citenamefont {Gaudin}, \citenamefont {Auffret}, \citenamefont {Rodmacq},
  \citenamefont {Schuhl}, \citenamefont {Pizzini}, \citenamefont {Vogel},\ and\
  \citenamefont {Gambardella}}]{miron2010current}%
  \BibitemOpen
  \bibfield  {author} {\bibinfo {author} {\bibfnamefont {I.~M.}\ \bibnamefont
  {Miron}}, \bibinfo {author} {\bibfnamefont {G.}~\bibnamefont {Gaudin}},
  \bibinfo {author} {\bibfnamefont {S.}~\bibnamefont {Auffret}}, \bibinfo
  {author} {\bibfnamefont {B.}~\bibnamefont {Rodmacq}}, \bibinfo {author}
  {\bibfnamefont {A.}~\bibnamefont {Schuhl}}, \bibinfo {author} {\bibfnamefont
  {S.}~\bibnamefont {Pizzini}}, \bibinfo {author} {\bibfnamefont
  {J.}~\bibnamefont {Vogel}}, \ and\ \bibinfo {author} {\bibfnamefont
  {P.}~\bibnamefont {Gambardella}},\ }\href@noop {} {\bibfield  {journal}
  {\bibinfo  {journal} {Nat. Mater.}\ }\textbf {\bibinfo {volume} {9}},\
  \bibinfo {pages} {230} (\bibinfo {year} {2010})}\BibitemShut {NoStop}%
\bibitem [{\citenamefont {Pi}\ \emph {et~al.}(2010)\citenamefont {Pi},
  \citenamefont {Won~Kim}, \citenamefont {Bae}, \citenamefont {Lee},
  \citenamefont {Cho}, \citenamefont {Kim},\ and\ \citenamefont
  {Seo}}]{pi2010tilting}%
  \BibitemOpen
  \bibfield  {author} {\bibinfo {author} {\bibfnamefont {U.~H.}\ \bibnamefont
  {Pi}}, \bibinfo {author} {\bibfnamefont {K.}~\bibnamefont {Won~Kim}},
  \bibinfo {author} {\bibfnamefont {J.~Y.}\ \bibnamefont {Bae}}, \bibinfo
  {author} {\bibfnamefont {S.~C.}\ \bibnamefont {Lee}}, \bibinfo {author}
  {\bibfnamefont {Y.~J.}\ \bibnamefont {Cho}}, \bibinfo {author} {\bibfnamefont
  {K.~S.}\ \bibnamefont {Kim}}, \ and\ \bibinfo {author} {\bibfnamefont
  {S.}~\bibnamefont {Seo}},\ }\href@noop {} {\bibfield  {journal} {\bibinfo
  {journal} {Appl. Phys. Lett.}\ }\textbf {\bibinfo {volume} {97}},\ \bibinfo
  {pages} {162507} (\bibinfo {year} {2010})}\BibitemShut {NoStop}%
\bibitem [{\citenamefont {Liu}\ \emph {et~al.}(2011)\citenamefont {Liu},
  \citenamefont {Moriyama}, \citenamefont {Ralph},\ and\ \citenamefont
  {Buhrman}}]{PhysRevLett.106.036601}%
  \BibitemOpen
  \bibfield  {author} {\bibinfo {author} {\bibfnamefont {L.}~\bibnamefont
  {Liu}}, \bibinfo {author} {\bibfnamefont {T.}~\bibnamefont {Moriyama}},
  \bibinfo {author} {\bibfnamefont {D.~C.}\ \bibnamefont {Ralph}}, \ and\
  \bibinfo {author} {\bibfnamefont {R.~A.}\ \bibnamefont {Buhrman}},\ }\href
  {\doibase 10.1103/PhysRevLett.106.036601} {\bibfield  {journal} {\bibinfo
  {journal} {Phys. Rev. Lett.}\ }\textbf {\bibinfo {volume} {106}},\ \bibinfo
  {pages} {036601} (\bibinfo {year} {2011})}\BibitemShut {NoStop}%
\bibitem [{\citenamefont {Miron}\ \emph {et~al.}(2011)\citenamefont {Miron},
  \citenamefont {Garello}, \citenamefont {Gaudin}, \citenamefont {Zermatten},
  \citenamefont {Costache}, \citenamefont {Auffret}, \citenamefont {Bandiera},
  \citenamefont {Rodmacq}, \citenamefont {Schuhl},\ and\ \citenamefont
  {Gambardella}}]{Miron2011}%
  \BibitemOpen
  \bibfield  {author} {\bibinfo {author} {\bibfnamefont {I.~M.}\ \bibnamefont
  {Miron}}, \bibinfo {author} {\bibfnamefont {K.}~\bibnamefont {Garello}},
  \bibinfo {author} {\bibfnamefont {G.}~\bibnamefont {Gaudin}}, \bibinfo
  {author} {\bibfnamefont {P.-J.}\ \bibnamefont {Zermatten}}, \bibinfo {author}
  {\bibfnamefont {M.~V.}\ \bibnamefont {Costache}}, \bibinfo {author}
  {\bibfnamefont {S.}~\bibnamefont {Auffret}}, \bibinfo {author} {\bibfnamefont
  {S.}~\bibnamefont {Bandiera}}, \bibinfo {author} {\bibfnamefont
  {B.}~\bibnamefont {Rodmacq}}, \bibinfo {author} {\bibfnamefont
  {A.}~\bibnamefont {Schuhl}}, \ and\ \bibinfo {author} {\bibfnamefont
  {P.}~\bibnamefont {Gambardella}},\ }\href@noop {} {\bibfield  {journal}
  {\bibinfo  {journal} {Nature}\ }\textbf {\bibinfo {volume} {476}},\ \bibinfo
  {pages} {189} (\bibinfo {year} {2011})}\BibitemShut {NoStop}%
\bibitem [{\citenamefont {Wang}\ \emph {et~al.}(2011)\citenamefont {Wang},
  \citenamefont {Sun}, \citenamefont {Wu}, \citenamefont {Tiberkevich},\ and\
  \citenamefont {Slavin}}]{PhysRevLett.107.146602}%
  \BibitemOpen
  \bibfield  {author} {\bibinfo {author} {\bibfnamefont {Z.}~\bibnamefont
  {Wang}}, \bibinfo {author} {\bibfnamefont {Y.}~\bibnamefont {Sun}}, \bibinfo
  {author} {\bibfnamefont {M.}~\bibnamefont {Wu}}, \bibinfo {author}
  {\bibfnamefont {V.}~\bibnamefont {Tiberkevich}}, \ and\ \bibinfo {author}
  {\bibfnamefont {A.}~\bibnamefont {Slavin}},\ }\href {\doibase
  10.1103/PhysRevLett.107.146602} {\bibfield  {journal} {\bibinfo  {journal}
  {Phys. Rev. Lett.}\ }\textbf {\bibinfo {volume} {107}},\ \bibinfo {pages}
  {146602} (\bibinfo {year} {2011})}\BibitemShut {NoStop}%
\bibitem [{\citenamefont {Liu}\ \emph {et~al.}(2012{\natexlab{a}})\citenamefont
  {Liu}, \citenamefont {Lee}, \citenamefont {Gudmundsen}, \citenamefont
  {Ralph},\ and\ \citenamefont {Buhrman}}]{PhysRevLett.109.096602}%
  \BibitemOpen
  \bibfield  {author} {\bibinfo {author} {\bibfnamefont {L.}~\bibnamefont
  {Liu}}, \bibinfo {author} {\bibfnamefont {O.~J.}\ \bibnamefont {Lee}},
  \bibinfo {author} {\bibfnamefont {T.~J.}\ \bibnamefont {Gudmundsen}},
  \bibinfo {author} {\bibfnamefont {D.~C.}\ \bibnamefont {Ralph}}, \ and\
  \bibinfo {author} {\bibfnamefont {R.~A.}\ \bibnamefont {Buhrman}},\ }\href
  {\doibase 10.1103/PhysRevLett.109.096602} {\bibfield  {journal} {\bibinfo
  {journal} {Phys. Rev. Lett.}\ }\textbf {\bibinfo {volume} {109}},\ \bibinfo
  {pages} {096602} (\bibinfo {year} {2012}{\natexlab{a}})}\BibitemShut
  {NoStop}%
\bibitem [{\citenamefont {Liu}\ \emph {et~al.}(2012{\natexlab{b}})\citenamefont
  {Liu}, \citenamefont {Pai}, \citenamefont {Li}, \citenamefont {Tseng},
  \citenamefont {Ralph},\ and\ \citenamefont {Buhrman}}]{liu2012spin}%
  \BibitemOpen
  \bibfield  {author} {\bibinfo {author} {\bibfnamefont {L.}~\bibnamefont
  {Liu}}, \bibinfo {author} {\bibfnamefont {C.-F.}\ \bibnamefont {Pai}},
  \bibinfo {author} {\bibfnamefont {Y.}~\bibnamefont {Li}}, \bibinfo {author}
  {\bibfnamefont {H.}~\bibnamefont {Tseng}}, \bibinfo {author} {\bibfnamefont
  {D.}~\bibnamefont {Ralph}}, \ and\ \bibinfo {author} {\bibfnamefont
  {R.}~\bibnamefont {Buhrman}},\ }\href@noop {} {\bibfield  {journal} {\bibinfo
   {journal} {Science}\ }\textbf {\bibinfo {volume} {336}},\ \bibinfo {pages}
  {555} (\bibinfo {year} {2012}{\natexlab{b}})}\BibitemShut {NoStop}%
\bibitem [{\citenamefont {Pai}\ \emph {et~al.}(2012)\citenamefont {Pai},
  \citenamefont {Liu}, \citenamefont {Li}, \citenamefont {Tseng}, \citenamefont
  {Ralph},\ and\ \citenamefont {Buhrman}}]{pai2012spin}%
  \BibitemOpen
  \bibfield  {author} {\bibinfo {author} {\bibfnamefont {C.-F.}\ \bibnamefont
  {Pai}}, \bibinfo {author} {\bibfnamefont {L.}~\bibnamefont {Liu}}, \bibinfo
  {author} {\bibfnamefont {Y.}~\bibnamefont {Li}}, \bibinfo {author}
  {\bibfnamefont {H.}~\bibnamefont {Tseng}}, \bibinfo {author} {\bibfnamefont
  {D.}~\bibnamefont {Ralph}}, \ and\ \bibinfo {author} {\bibfnamefont
  {R.}~\bibnamefont {Buhrman}},\ }\href@noop {} {\bibfield  {journal} {\bibinfo
   {journal} {Appl. Phys. Lett.}\ }\textbf {\bibinfo {volume} {101}},\ \bibinfo
  {pages} {122404} (\bibinfo {year} {2012})}\BibitemShut {NoStop}%
\bibitem [{\citenamefont {Kim}\ \emph {et~al.}(2012{\natexlab{a}})\citenamefont
  {Kim}, \citenamefont {Sinha}, \citenamefont {Hayashi}, \citenamefont
  {Yamanouchi}, \citenamefont {Fukami}, \citenamefont {Suzuki}, \citenamefont
  {Mitani},\ and\ \citenamefont {Ohno}}]{kim2012layer}%
  \BibitemOpen
  \bibfield  {author} {\bibinfo {author} {\bibfnamefont {J.}~\bibnamefont
  {Kim}}, \bibinfo {author} {\bibfnamefont {J.}~\bibnamefont {Sinha}}, \bibinfo
  {author} {\bibfnamefont {M.}~\bibnamefont {Hayashi}}, \bibinfo {author}
  {\bibfnamefont {M.}~\bibnamefont {Yamanouchi}}, \bibinfo {author}
  {\bibfnamefont {S.}~\bibnamefont {Fukami}}, \bibinfo {author} {\bibfnamefont
  {T.}~\bibnamefont {Suzuki}}, \bibinfo {author} {\bibfnamefont
  {S.}~\bibnamefont {Mitani}}, \ and\ \bibinfo {author} {\bibfnamefont
  {H.}~\bibnamefont {Ohno}},\ }\href@noop {} {\bibfield  {journal} {\bibinfo
  {journal} {Nat. Mater.}\ } (\bibinfo {year}
  {2012}{\natexlab{a}})}\BibitemShut {NoStop}%
\bibitem [{\citenamefont {Bychkov}\ and\ \citenamefont
  {Rashba}(1984)}]{bychkov1984oscillatory}%
  \BibitemOpen
  \bibfield  {author} {\bibinfo {author} {\bibfnamefont {Y.~A.}\ \bibnamefont
  {Bychkov}}\ and\ \bibinfo {author} {\bibfnamefont {E.~I.}\ \bibnamefont
  {Rashba}},\ }\href@noop {} {\bibfield  {journal} {\bibinfo  {journal}
  {Journal of physics C: Solid state physics}\ }\textbf {\bibinfo {volume}
  {17}},\ \bibinfo {pages} {6039} (\bibinfo {year} {1984})}\BibitemShut
  {NoStop}%
\bibitem [{\citenamefont {Manchon}\ and\ \citenamefont
  {Zhang}(2008)}]{PhysRevB.78.212405}%
  \BibitemOpen
  \bibfield  {author} {\bibinfo {author} {\bibfnamefont {A.}~\bibnamefont
  {Manchon}}\ and\ \bibinfo {author} {\bibfnamefont {S.}~\bibnamefont
  {Zhang}},\ }\href {\doibase 10.1103/PhysRevB.78.212405} {\bibfield  {journal}
  {\bibinfo  {journal} {Phys. Rev. B}\ }\textbf {\bibinfo {volume} {78}},\
  \bibinfo {pages} {212405} (\bibinfo {year} {2008})}\BibitemShut {NoStop}%
\bibitem [{\citenamefont {Manchon}\ and\ \citenamefont
  {Zhang}(2009)}]{PhysRevB.79.094422}%
  \BibitemOpen
  \bibfield  {author} {\bibinfo {author} {\bibfnamefont {A.}~\bibnamefont
  {Manchon}}\ and\ \bibinfo {author} {\bibfnamefont {S.}~\bibnamefont
  {Zhang}},\ }\href {\doibase 10.1103/PhysRevB.79.094422} {\bibfield  {journal}
  {\bibinfo  {journal} {Phys. Rev. B}\ }\textbf {\bibinfo {volume} {79}},\
  \bibinfo {pages} {094422} (\bibinfo {year} {2009})}\BibitemShut {NoStop}%
\bibitem [{\citenamefont {Matos-Abiague}\ and\ \citenamefont
  {Rodriguez-Suarez}(2009)}]{matos2009spin}%
  \BibitemOpen
  \bibfield  {author} {\bibinfo {author} {\bibfnamefont {A.}~\bibnamefont
  {Matos-Abiague}}\ and\ \bibinfo {author} {\bibfnamefont {R.}~\bibnamefont
  {Rodriguez-Suarez}},\ }\href@noop {} {\bibfield  {journal} {\bibinfo
  {journal} {Phys. Rev. B}\ }\textbf {\bibinfo {volume} {80}},\ \bibinfo
  {pages} {094424} (\bibinfo {year} {2009})}\BibitemShut {NoStop}%
\bibitem [{\citenamefont {Gambardella}\ and\ \citenamefont
  {Miron}(2011)}]{gambardella2011current}%
  \BibitemOpen
  \bibfield  {author} {\bibinfo {author} {\bibfnamefont {P.}~\bibnamefont
  {Gambardella}}\ and\ \bibinfo {author} {\bibfnamefont {I.~M.}\ \bibnamefont
  {Miron}},\ }\href@noop {} {\bibfield  {journal} {\bibinfo  {journal}
  {Philosophical Transactions of the Royal Society A: Mathematical, Physical
  and Engineering Sciences}\ }\textbf {\bibinfo {volume} {369}},\ \bibinfo
  {pages} {3175} (\bibinfo {year} {2011})}\BibitemShut {NoStop}%
\bibitem [{\citenamefont {Haney}\ \emph
  {et~al.}(2013{\natexlab{a}})\citenamefont {Haney}, \citenamefont {Lee},
  \citenamefont {Lee}, \citenamefont {Manchon},\ and\ \citenamefont
  {Stiles}}]{PhysRevB.88.214417}%
  \BibitemOpen
  \bibfield  {author} {\bibinfo {author} {\bibfnamefont {P.~M.}\ \bibnamefont
  {Haney}}, \bibinfo {author} {\bibfnamefont {H.-W.}\ \bibnamefont {Lee}},
  \bibinfo {author} {\bibfnamefont {K.-J.}\ \bibnamefont {Lee}}, \bibinfo
  {author} {\bibfnamefont {A.}~\bibnamefont {Manchon}}, \ and\ \bibinfo
  {author} {\bibfnamefont {M.~D.}\ \bibnamefont {Stiles}},\ }\href {\doibase
  10.1103/PhysRevB.88.214417} {\bibfield  {journal} {\bibinfo  {journal} {Phys.
  Rev. B}\ }\textbf {\bibinfo {volume} {88}},\ \bibinfo {pages} {214417}
  (\bibinfo {year} {2013}{\natexlab{a}})}\BibitemShut {NoStop}%
\bibitem [{\citenamefont {Hirsch}(1999)}]{PhysRevLett.83.1834}%
  \BibitemOpen
  \bibfield  {author} {\bibinfo {author} {\bibfnamefont {J.~E.}\ \bibnamefont
  {Hirsch}},\ }\href {\doibase 10.1103/PhysRevLett.83.1834} {\bibfield
  {journal} {\bibinfo  {journal} {Phys. Rev. Lett.}\ }\textbf {\bibinfo
  {volume} {83}},\ \bibinfo {pages} {1834} (\bibinfo {year}
  {1999})}\BibitemShut {NoStop}%
\bibitem [{\citenamefont {Zhang}(2000)}]{PhysRevLett.85.393}%
  \BibitemOpen
  \bibfield  {author} {\bibinfo {author} {\bibfnamefont {S.}~\bibnamefont
  {Zhang}},\ }\href {\doibase 10.1103/PhysRevLett.85.393} {\bibfield  {journal}
  {\bibinfo  {journal} {Phys. Rev. Lett.}\ }\textbf {\bibinfo {volume} {85}},\
  \bibinfo {pages} {393} (\bibinfo {year} {2000})}\BibitemShut {NoStop}%
\bibitem [{\citenamefont {Sinova}\ \emph {et~al.}(2004)\citenamefont {Sinova},
  \citenamefont {Culcer}, \citenamefont {Niu}, \citenamefont {Sinitsyn},
  \citenamefont {Jungwirth},\ and\ \citenamefont
  {MacDonald}}]{PhysRevLett.92.126603}%
  \BibitemOpen
  \bibfield  {author} {\bibinfo {author} {\bibfnamefont {J.}~\bibnamefont
  {Sinova}}, \bibinfo {author} {\bibfnamefont {D.}~\bibnamefont {Culcer}},
  \bibinfo {author} {\bibfnamefont {Q.}~\bibnamefont {Niu}}, \bibinfo {author}
  {\bibfnamefont {N.~A.}\ \bibnamefont {Sinitsyn}}, \bibinfo {author}
  {\bibfnamefont {T.}~\bibnamefont {Jungwirth}}, \ and\ \bibinfo {author}
  {\bibfnamefont {A.~H.}\ \bibnamefont {MacDonald}},\ }\href {\doibase
  10.1103/PhysRevLett.92.126603} {\bibfield  {journal} {\bibinfo  {journal}
  {Phys. Rev. Lett.}\ }\textbf {\bibinfo {volume} {92}},\ \bibinfo {pages}
  {126603} (\bibinfo {year} {2004})}\BibitemShut {NoStop}%
\bibitem [{\citenamefont {Jungwirth}\ \emph {et~al.}(2012)\citenamefont
  {Jungwirth}, \citenamefont {Wunderlich},\ and\ \citenamefont
  {Olejn{\'\i}k}}]{jungwirth2012spin}%
  \BibitemOpen
  \bibfield  {author} {\bibinfo {author} {\bibfnamefont {T.}~\bibnamefont
  {Jungwirth}}, \bibinfo {author} {\bibfnamefont {J.}~\bibnamefont
  {Wunderlich}}, \ and\ \bibinfo {author} {\bibfnamefont {K.}~\bibnamefont
  {Olejn{\'\i}k}},\ }\href@noop {} {\bibfield  {journal} {\bibinfo  {journal}
  {Nat. Mater.}\ }\textbf {\bibinfo {volume} {11}},\ \bibinfo {pages} {382}
  (\bibinfo {year} {2012})}\BibitemShut {NoStop}%
\bibitem [{\citenamefont {Slonczewski}(1996)}]{slonczewski1996current}%
  \BibitemOpen
  \bibfield  {author} {\bibinfo {author} {\bibfnamefont {J.~C.}\ \bibnamefont
  {Slonczewski}},\ }\href@noop {} {\bibfield  {journal} {\bibinfo  {journal}
  {J. Magn. Magn. Mater.}\ }\textbf {\bibinfo {volume} {159}},\ \bibinfo
  {pages} {L1} (\bibinfo {year} {1996})}\BibitemShut {NoStop}%
\bibitem [{\citenamefont {Ralph}\ and\ \citenamefont
  {Stiles}(2008)}]{ralph2008spin}%
  \BibitemOpen
  \bibfield  {author} {\bibinfo {author} {\bibfnamefont {D.}~\bibnamefont
  {Ralph}}\ and\ \bibinfo {author} {\bibfnamefont {M.~D.}\ \bibnamefont
  {Stiles}},\ }\href@noop {} {\bibfield  {journal} {\bibinfo  {journal} {J.
  Magn. Magn. Mater.}\ }\textbf {\bibinfo {volume} {320}},\ \bibinfo {pages}
  {1190} (\bibinfo {year} {2008})}\BibitemShut {NoStop}%
\bibitem [{\citenamefont {Haney}\ \emph
  {et~al.}(2013{\natexlab{b}})\citenamefont {Haney}, \citenamefont {Lee},
  \citenamefont {Lee}, \citenamefont {Manchon},\ and\ \citenamefont
  {Stiles}}]{PhysRevB.87.174411}%
  \BibitemOpen
  \bibfield  {author} {\bibinfo {author} {\bibfnamefont {P.~M.}\ \bibnamefont
  {Haney}}, \bibinfo {author} {\bibfnamefont {H.-W.}\ \bibnamefont {Lee}},
  \bibinfo {author} {\bibfnamefont {K.-J.}\ \bibnamefont {Lee}}, \bibinfo
  {author} {\bibfnamefont {A.}~\bibnamefont {Manchon}}, \ and\ \bibinfo
  {author} {\bibfnamefont {M.~D.}\ \bibnamefont {Stiles}},\ }\href {\doibase
  10.1103/PhysRevB.87.174411} {\bibfield  {journal} {\bibinfo  {journal} {Phys.
  Rev. B}\ }\textbf {\bibinfo {volume} {87}},\ \bibinfo {pages} {174411}
  (\bibinfo {year} {2013}{\natexlab{b}})}\BibitemShut {NoStop}%
\bibitem [{\citenamefont {Yu}\ \emph {et~al.}(2014)\citenamefont {Yu},
  \citenamefont {Upadhyaya}, \citenamefont {Wong}, \citenamefont {Jiang},
  \citenamefont {Alzate}, \citenamefont {Tang}, \citenamefont {Amiri},\ and\
  \citenamefont {Wang}}]{PhysRevB.89.104421}%
  \BibitemOpen
  \bibfield  {author} {\bibinfo {author} {\bibfnamefont {G.}~\bibnamefont
  {Yu}}, \bibinfo {author} {\bibfnamefont {P.}~\bibnamefont {Upadhyaya}},
  \bibinfo {author} {\bibfnamefont {K.~L.}\ \bibnamefont {Wong}}, \bibinfo
  {author} {\bibfnamefont {W.}~\bibnamefont {Jiang}}, \bibinfo {author}
  {\bibfnamefont {J.~G.}\ \bibnamefont {Alzate}}, \bibinfo {author}
  {\bibfnamefont {J.}~\bibnamefont {Tang}}, \bibinfo {author} {\bibfnamefont
  {P.~K.}\ \bibnamefont {Amiri}}, \ and\ \bibinfo {author} {\bibfnamefont
  {K.~L.}\ \bibnamefont {Wang}},\ }\href {\doibase 10.1103/PhysRevB.89.104421}
  {\bibfield  {journal} {\bibinfo  {journal} {Phys. Rev. B}\ }\textbf {\bibinfo
  {volume} {89}},\ \bibinfo {pages} {104421} (\bibinfo {year}
  {2014})}\BibitemShut {NoStop}%
\bibitem [{\citenamefont {Kim}\ \emph {et~al.}(2012{\natexlab{b}})\citenamefont
  {Kim}, \citenamefont {Seo}, \citenamefont {Ryu}, \citenamefont {Lee},\ and\
  \citenamefont {Lee}}]{PhysRevB.85.180404}%
  \BibitemOpen
  \bibfield  {author} {\bibinfo {author} {\bibfnamefont {K.-W.}\ \bibnamefont
  {Kim}}, \bibinfo {author} {\bibfnamefont {S.-M.}\ \bibnamefont {Seo}},
  \bibinfo {author} {\bibfnamefont {J.}~\bibnamefont {Ryu}}, \bibinfo {author}
  {\bibfnamefont {K.-J.}\ \bibnamefont {Lee}}, \ and\ \bibinfo {author}
  {\bibfnamefont {H.-W.}\ \bibnamefont {Lee}},\ }\href {\doibase
  10.1103/PhysRevB.85.180404} {\bibfield  {journal} {\bibinfo  {journal} {Phys.
  Rev. B}\ }\textbf {\bibinfo {volume} {85}},\ \bibinfo {pages} {180404}
  (\bibinfo {year} {2012}{\natexlab{b}})}\BibitemShut {NoStop}%
\bibitem [{\citenamefont {Pesin}\ and\ \citenamefont
  {MacDonald}(2012)}]{PhysRevB.86.014416}%
  \BibitemOpen
  \bibfield  {author} {\bibinfo {author} {\bibfnamefont {D.~A.}\ \bibnamefont
  {Pesin}}\ and\ \bibinfo {author} {\bibfnamefont {A.~H.}\ \bibnamefont
  {MacDonald}},\ }\href {\doibase 10.1103/PhysRevB.86.014416} {\bibfield
  {journal} {\bibinfo  {journal} {Phys. Rev. B}\ }\textbf {\bibinfo {volume}
  {86}},\ \bibinfo {pages} {014416} (\bibinfo {year} {2012})}\BibitemShut
  {NoStop}%
\bibitem [{\citenamefont {Wang}\ and\ \citenamefont
  {Manchon}(2012)}]{PhysRevLett.108.117201}%
  \BibitemOpen
  \bibfield  {author} {\bibinfo {author} {\bibfnamefont {X.}~\bibnamefont
  {Wang}}\ and\ \bibinfo {author} {\bibfnamefont {A.}~\bibnamefont {Manchon}},\
  }\href {\doibase 10.1103/PhysRevLett.108.117201} {\bibfield  {journal}
  {\bibinfo  {journal} {Phys. Rev. Lett.}\ }\textbf {\bibinfo {volume} {108}},\
  \bibinfo {pages} {117201} (\bibinfo {year} {2012})}\BibitemShut {NoStop}%
\bibitem [{\citenamefont {Stoner}\ and\ \citenamefont
  {Wohlfarth}(1948)}]{stoner1948mechanism}%
  \BibitemOpen
  \bibfield  {author} {\bibinfo {author} {\bibfnamefont {E.~C.}\ \bibnamefont
  {Stoner}}\ and\ \bibinfo {author} {\bibfnamefont {E.}~\bibnamefont
  {Wohlfarth}},\ }\href@noop {} {\bibfield  {journal} {\bibinfo  {journal}
  {Philosophical Transactions of the Royal Society of London. Series A.
  Mathematical and Physical Sciences}\ ,\ \bibinfo {pages} {599}} (\bibinfo
  {year} {1948})}\BibitemShut {NoStop}%
\bibitem [{\citenamefont {Thiaville}(2000)}]{PhysRevB.61.12221}%
  \BibitemOpen
  \bibfield  {author} {\bibinfo {author} {\bibfnamefont {A.}~\bibnamefont
  {Thiaville}},\ }\href {\doibase 10.1103/PhysRevB.61.12221} {\bibfield
  {journal} {\bibinfo  {journal} {Phys. Rev. B}\ }\textbf {\bibinfo {volume}
  {61}},\ \bibinfo {pages} {12221} (\bibinfo {year} {2000})}\BibitemShut
  {NoStop}%
\bibitem [{\citenamefont {Yan}\ \emph {et~al.}(2013)\citenamefont {Yan},
  \citenamefont {Sun},\ and\ \citenamefont {Bazaliy}}]{PhysRevB.88.054408}%
  \BibitemOpen
  \bibfield  {author} {\bibinfo {author} {\bibfnamefont {S.}~\bibnamefont
  {Yan}}, \bibinfo {author} {\bibfnamefont {Z.}~\bibnamefont {Sun}}, \ and\
  \bibinfo {author} {\bibfnamefont {Y.~B.}\ \bibnamefont {Bazaliy}},\ }\href
  {\doibase 10.1103/PhysRevB.88.054408} {\bibfield  {journal} {\bibinfo
  {journal} {Phys. Rev. B}\ }\textbf {\bibinfo {volume} {88}},\ \bibinfo
  {pages} {054408} (\bibinfo {year} {2013})}\BibitemShut {NoStop}%
\bibitem [{\citenamefont {Chang}\ \emph
  {et~al.}(2011{\natexlab{a}})\citenamefont {Chang}, \citenamefont {Chen},\
  and\ \citenamefont {Chang}}]{PhysRevB.83.054425}%
  \BibitemOpen
  \bibfield  {author} {\bibinfo {author} {\bibfnamefont {J.~H.}\ \bibnamefont
  {Chang}}, \bibinfo {author} {\bibfnamefont {H.~H.}\ \bibnamefont {Chen}}, \
  and\ \bibinfo {author} {\bibfnamefont {C.~R.}\ \bibnamefont {Chang}},\ }\href
  {\doibase 10.1103/PhysRevB.83.054425} {\bibfield  {journal} {\bibinfo
  {journal} {Phys. Rev. B}\ }\textbf {\bibinfo {volume} {83}},\ \bibinfo
  {pages} {054425} (\bibinfo {year} {2011}{\natexlab{a}})}\BibitemShut
  {NoStop}%
\bibitem [{\citenamefont {Lee}\ \emph {et~al.}(2013)\citenamefont {Lee},
  \citenamefont {Lee}, \citenamefont {Min},\ and\ \citenamefont
  {Lee}}]{lee2013threshold}%
  \BibitemOpen
  \bibfield  {author} {\bibinfo {author} {\bibfnamefont {K.-S.}\ \bibnamefont
  {Lee}}, \bibinfo {author} {\bibfnamefont {S.-W.}\ \bibnamefont {Lee}},
  \bibinfo {author} {\bibfnamefont {B.-C.}\ \bibnamefont {Min}}, \ and\
  \bibinfo {author} {\bibfnamefont {K.-J.}\ \bibnamefont {Lee}},\ }\href@noop
  {} {\bibfield  {journal} {\bibinfo  {journal} {Appl. Phys. Lett.}\ }\textbf
  {\bibinfo {volume} {102}},\ \bibinfo {pages} {112410} (\bibinfo {year}
  {2013})}\BibitemShut {NoStop}%
\bibitem [{\citenamefont {Hirsch}\ \emph {et~al.}(2004)\citenamefont {Hirsch},
  \citenamefont {Smale},\ and\ \citenamefont
  {Devaney}}]{hirsch2004differential}%
  \BibitemOpen
  \bibfield  {author} {\bibinfo {author} {\bibfnamefont {M.~W.}\ \bibnamefont
  {Hirsch}}, \bibinfo {author} {\bibfnamefont {S.}~\bibnamefont {Smale}}, \
  and\ \bibinfo {author} {\bibfnamefont {R.~L.}\ \bibnamefont {Devaney}},\
  }\href@noop {} {\emph {\bibinfo {title} {Differential equations, dynamical
  systems, and an introduction to chaos}}},\ Vol.~\bibinfo {volume} {60}\
  (\bibinfo  {publisher} {Academic press},\ \bibinfo {year} {2004})\BibitemShut
  {NoStop}%
\bibitem [{\citenamefont {Sodemann}\ and\ \citenamefont
  {Bazaliy}(2011)}]{PhysRevB.84.064422}%
  \BibitemOpen
  \bibfield  {author} {\bibinfo {author} {\bibfnamefont {I.}~\bibnamefont
  {Sodemann}}\ and\ \bibinfo {author} {\bibfnamefont {Y.~B.}\ \bibnamefont
  {Bazaliy}},\ }\href {\doibase 10.1103/PhysRevB.84.064422} {\bibfield
  {journal} {\bibinfo  {journal} {Phys. Rev. B}\ }\textbf {\bibinfo {volume}
  {84}},\ \bibinfo {pages} {064422} (\bibinfo {year} {2011})}\BibitemShut
  {NoStop}%
\bibitem [{\citenamefont {Chang}\ \emph
  {et~al.}(2011{\natexlab{b}})\citenamefont {Chang}, \citenamefont {Chen},\
  and\ \citenamefont {Chang}}]{chang2011multiple}%
  \BibitemOpen
  \bibfield  {author} {\bibinfo {author} {\bibfnamefont {J.-H.}\ \bibnamefont
  {Chang}}, \bibinfo {author} {\bibfnamefont {H.-H.}\ \bibnamefont {Chen}}, \
  and\ \bibinfo {author} {\bibfnamefont {C.-R.}\ \bibnamefont {Chang}},\
  }\href@noop {} {\bibfield  {journal} {\bibinfo  {journal} {Magnetics, IEEE
  Transactions on}\ }\textbf {\bibinfo {volume} {47}},\ \bibinfo {pages} {3876}
  (\bibinfo {year} {2011}{\natexlab{b}})}\BibitemShut {NoStop}%
\bibitem [{\citenamefont {Bertotti}\ \emph {et~al.}(2003)\citenamefont
  {Bertotti}, \citenamefont {Mayergoyz}, \citenamefont {Serpico},\ and\
  \citenamefont {Dimian}}]{bertotti2003comparison}%
  \BibitemOpen
  \bibfield  {author} {\bibinfo {author} {\bibfnamefont {G.}~\bibnamefont
  {Bertotti}}, \bibinfo {author} {\bibfnamefont {I.}~\bibnamefont {Mayergoyz}},
  \bibinfo {author} {\bibfnamefont {C.}~\bibnamefont {Serpico}}, \ and\
  \bibinfo {author} {\bibfnamefont {M.}~\bibnamefont {Dimian}},\ }\href@noop {}
  {\bibfield  {journal} {\bibinfo  {journal} {J. Appl. Phys.}\ }\textbf
  {\bibinfo {volume} {93}},\ \bibinfo {pages} {6811} (\bibinfo {year}
  {2003})}\BibitemShut {NoStop}%
\bibitem [{\citenamefont {Bazaliy}\ and\ \citenamefont
  {Stankiewicz}(2011)}]{bazaliy2011ballistic}%
  \BibitemOpen
  \bibfield  {author} {\bibinfo {author} {\bibfnamefont {Y.~B.}\ \bibnamefont
  {Bazaliy}}\ and\ \bibinfo {author} {\bibfnamefont {A.}~\bibnamefont
  {Stankiewicz}},\ }\href@noop {} {\bibfield  {journal} {\bibinfo  {journal}
  {Appl. Phys. Lett.}\ }\textbf {\bibinfo {volume} {98}},\ \bibinfo {pages}
  {142501} (\bibinfo {year} {2011})}\BibitemShut {NoStop}%
\bibitem [{\citenamefont {Bazaliy}(2011)}]{bazaliy2011analytic}%
  \BibitemOpen
  \bibfield  {author} {\bibinfo {author} {\bibfnamefont {Y.~B.}\ \bibnamefont
  {Bazaliy}},\ }\href@noop {} {\bibfield  {journal} {\bibinfo  {journal} {J.
  Appl. Phys.}\ }\textbf {\bibinfo {volume} {110}},\ \bibinfo {pages} {063920}
  (\bibinfo {year} {2011})}\BibitemShut {NoStop}%
\bibitem [{\citenamefont {Koch}\ \emph {et~al.}(2004)\citenamefont {Koch},
  \citenamefont {Katine},\ and\ \citenamefont {Sun}}]{koch2004time}%
  \BibitemOpen
  \bibfield  {author} {\bibinfo {author} {\bibfnamefont {R.}~\bibnamefont
  {Koch}}, \bibinfo {author} {\bibfnamefont {J.}~\bibnamefont {Katine}}, \ and\
  \bibinfo {author} {\bibfnamefont {J.}~\bibnamefont {Sun}},\ }\href@noop {}
  {\bibfield  {journal} {\bibinfo  {journal} {Physical review letters}\
  }\textbf {\bibinfo {volume} {92}},\ \bibinfo {pages} {088302} (\bibinfo
  {year} {2004})}\BibitemShut {NoStop}%
\bibitem [{\citenamefont {Sun}(2000)}]{sun2000spin}%
  \BibitemOpen
  \bibfield  {author} {\bibinfo {author} {\bibfnamefont {J.}~\bibnamefont
  {Sun}},\ }\href@noop {} {\bibfield  {journal} {\bibinfo  {journal} {Physical
  Review B}\ }\textbf {\bibinfo {volume} {62}},\ \bibinfo {pages} {570}
  (\bibinfo {year} {2000})}\BibitemShut {NoStop}%
\bibitem [{\citenamefont {Bauer}\ \emph {et~al.}(2000)\citenamefont {Bauer},
  \citenamefont {Fassbender}, \citenamefont {Hillebrands},\ and\ \citenamefont
  {Stamps}}]{bauer2000switching}%
  \BibitemOpen
  \bibfield  {author} {\bibinfo {author} {\bibfnamefont {M.}~\bibnamefont
  {Bauer}}, \bibinfo {author} {\bibfnamefont {J.}~\bibnamefont {Fassbender}},
  \bibinfo {author} {\bibfnamefont {B.}~\bibnamefont {Hillebrands}}, \ and\
  \bibinfo {author} {\bibfnamefont {R.}~\bibnamefont {Stamps}},\ }\href@noop {}
  {\bibfield  {journal} {\bibinfo  {journal} {Phys. Rev. B}\ }\textbf {\bibinfo
  {volume} {61}},\ \bibinfo {pages} {3410} (\bibinfo {year}
  {2000})}\BibitemShut {NoStop}%
\bibitem [{\citenamefont {Kent}\ \emph {et~al.}(2004)\citenamefont {Kent},
  \citenamefont {Ozyilmaz},\ and\ \citenamefont {Del~Barco}}]{kent2004spin}%
  \BibitemOpen
  \bibfield  {author} {\bibinfo {author} {\bibfnamefont {A.}~\bibnamefont
  {Kent}}, \bibinfo {author} {\bibfnamefont {B.}~\bibnamefont {Ozyilmaz}}, \
  and\ \bibinfo {author} {\bibfnamefont {E.}~\bibnamefont {Del~Barco}},\
  }\href@noop {} {\bibfield  {journal} {\bibinfo  {journal} {Appl. Phys.
  Lett.}\ }\textbf {\bibinfo {volume} {84}},\ \bibinfo {pages} {3897} (\bibinfo
  {year} {2004})}\BibitemShut {NoStop}%
\bibitem [{\citenamefont {Liu}\ \emph {et~al.}(2010)\citenamefont {Liu},
  \citenamefont {Bedau}, \citenamefont {Backes}, \citenamefont {Katine},
  \citenamefont {Langer},\ and\ \citenamefont {Kent}}]{liu2010ultrafast}%
  \BibitemOpen
  \bibfield  {author} {\bibinfo {author} {\bibfnamefont {H.}~\bibnamefont
  {Liu}}, \bibinfo {author} {\bibfnamefont {D.}~\bibnamefont {Bedau}}, \bibinfo
  {author} {\bibfnamefont {D.}~\bibnamefont {Backes}}, \bibinfo {author}
  {\bibfnamefont {J.}~\bibnamefont {Katine}}, \bibinfo {author} {\bibfnamefont
  {J.}~\bibnamefont {Langer}}, \ and\ \bibinfo {author} {\bibfnamefont
  {A.}~\bibnamefont {Kent}},\ }\href@noop {} {\bibfield  {journal} {\bibinfo
  {journal} {Appl. Phys. Lett.}\ }\textbf {\bibinfo {volume} {97}},\ \bibinfo
  {pages} {242510} (\bibinfo {year} {2010})}\BibitemShut {NoStop}%
\bibitem [{\citenamefont {Finocchio}\ \emph {et~al.}(2013)\citenamefont
  {Finocchio}, \citenamefont {Carpentieri}, \citenamefont {Martinez},\ and\
  \citenamefont {Azzerboni}}]{finocchio2013switching}%
  \BibitemOpen
  \bibfield  {author} {\bibinfo {author} {\bibfnamefont {G.}~\bibnamefont
  {Finocchio}}, \bibinfo {author} {\bibfnamefont {M.}~\bibnamefont
  {Carpentieri}}, \bibinfo {author} {\bibfnamefont {E.}~\bibnamefont
  {Martinez}}, \ and\ \bibinfo {author} {\bibfnamefont {B.}~\bibnamefont
  {Azzerboni}},\ }\href@noop {} {\bibfield  {journal} {\bibinfo  {journal}
  {Appl. Phys. Lett.}\ }\textbf {\bibinfo {volume} {102}},\ \bibinfo {pages}
  {212410} (\bibinfo {year} {2013})}\BibitemShut {NoStop}%
\bibitem [{\citenamefont {Yan}(2014)}]{yan2014nonlinear}%
  \BibitemOpen
  \bibfield  {author} {\bibinfo {author} {\bibfnamefont {S.}~\bibnamefont
  {Yan}},\ }\emph {\bibinfo {title} {Nonlinear Magnetic Dynamics and The
  Switching Phase Diagrams in Spintronic Devices}},\ \href@noop {} {Ph.D.
  thesis},\ \bibinfo  {school} {University of South Carolina} (\bibinfo {year}
  {2014})\BibitemShut {NoStop}%
\bibitem [{\citenamefont {d'Aquino}(2004)}]{d2004nonlinear}%
  \BibitemOpen
  \bibfield  {author} {\bibinfo {author} {\bibfnamefont {M.}~\bibnamefont
  {d'Aquino}},\ }\emph {\bibinfo {title} {Nonlinear magnetization dynamics in
  thin-films and nanoparticles}},\ \href@noop {} {Ph.D. thesis},\ \bibinfo
  {school} {PhD thesis, Universita degli studi di Napoli ``Federico II'',
  Facolta di Ingegneria, 2004. URL http://www. fedoa. unina. it/148} (\bibinfo
  {year} {2004})\BibitemShut {NoStop}%
\end{thebibliography}%

\end{document}